\documentclass[%
reprint,
amsmath,amssymb,
aps,
]{revtex4-1}

\usepackage{graphicx}
\usepackage{dcolumn}
\usepackage{bm}
\usepackage{hyperref}
\usepackage[mathlines]{lineno}
\usepackage[capitalise]{cleveref}
\usepackage{todonotes}
\usepackage{setspace}
\usepackage{physics}
\usepackage{amsmath}
\usepackage{amsthm}
\usepackage{svg}
\usepackage{mathtools}
\usepackage[autostyle]{csquotes}  
\usepackage{qcircuit}
\usepackage{tikz}
\usepackage{algorithm}
\usepackage{algpseudocode}
\usepackage{verbatim}
\usepackage{wasysym}
\usepackage[export]{adjustbox}

\hypersetup{colorlinks=true}

\DeclarePairedDelimiter\ceil{\lceil}{\rceil}

\begin{document}
	
	\preprint{APS/123-QED}
	
	\title{Combinatorial optimisation via highly efficient quantum walks}
	
	\date{\today}
	
	\author{S. Marsh}
	\email{samuel.marsh@research.uwa.edu.au}
	\affiliation{Department of Physics, The University of Western Australia, Perth, Australia}
	
	\author{J. B. Wang}
	\email{jingbo.wang@uwa.edu.au}
	\affiliation{Department of Physics, The University of Western Australia, Perth, Australia}
	
	\begin{abstract}
		We present a highly efficient quantum circuit for performing continuous time quantum walks (CTQWs) over an exponentially large set of combinatorial objects, provided that the objects can be indexed efficiently. CTQWs form the core mixing operation of a generalised version of the Quantum Approximate Optimisation Algorithm, which works by `steering' the quantum amplitude into high-quality solutions. The efficient quantum circuit holds the promise of finding high-quality solutions to certain classes of NP-hard combinatorial problems such as the Travelling Salesman Problem, maximum set splitting, graph partitioning, and lattice path optimisation.
	\end{abstract}
	
	
	\maketitle
	
	\section{Introduction}
	
	
	Combinatorial optimisation problems are known to be notoriously difficult to solve, even approximately in general~\cite{Korte:2007}.
	Quantum algorithms are able to solve these problems more efficiently, with a brute force quantum search offering a guaranteed square root speedup over the classical approach \cite{Grover1999, brassard2000}. Such speed-up is unfortunately insufficient to provide practically useful solutions, since these combinatorial optimisation problems scale up exponentially.
	
	\citet{Farhi2014} proposed the Quantum Approximate Optimisation Algorithm (QAOA), derived from approximating the quantum adiabatic algorithm on a gate model quantum computer, to find high quality solutions for general combinatorial optimisation problems~\cite{Farhi2014}. More recently, we extended the QAOA algorithm to solve constrained combinatorial optimisation problems via alternating continuous-time quantum walks over efficiently identifiable feasible solutions and solution-quality-dependent phase shifts~\cite{Marsh2019}. Throughout this paper, we refer to this quantum-walk-assisted generalisation as QWOA.
	
	
	
	The core component of QWOA is the continuous time quantum walk (CTQW) \cite{Farhi1998,Kempe2003,Childs2013,QWBook}, which acts as a `mixing' operator for the algorithm, with probability amplitudes transferred between feasible solutions of the problem. 
	A CTQW over the undirected graph with adjacency matrix $A$ is defined by the propagator $\hat{U}(t) = e^{-i t A}$. 
	Quantum walks have markedly different behaviour to classical random walks due to intrinsic quantum correlations and interference \cite{Tang2018,Spatialsearch,Engel2007,Berry2010}, and they have played a central role in quantum simulation and quantum information processing \cite{Lloyd1996,ambainis2007quantum,Douglas2008,Berry2011,Schreiber2012,Izaac2017a,Fleet2017,Ming2019,Tai2017,Harris2017,Yan753,Mirieaar7709}. CTQWs are particularly well-known for their applications to quantum spatial search \cite{Spatialsearch,Morley2019,Callison2019}, where the system is evolved for a sufficient length of time under the addition of the graph Hamiltonian $A$ and an oracular Hamiltonian encoding the marked element(s). However in QWOA we apply CTQWs independently, where a quantum circuit for $\hat{U}(t) = e^{-i t A}$ is used to map some initial amplitude distribution over the vertices to the distribution obtained after `walking' for time $t$. The oracular Hamiltonian encoding solution qualities is then applied sequentially, interleaved with further CTQWs.
	Of significance is that, for graph structures considered in this paper, the runtime of a QWOA circuit can be made independent of the walk times, leading to a distinct algorithmic advantage.
	
	In this paper, we discuss a significant and innovative application of QWOA to a wide range of combinatorial domains, which is defined as the set of feasible solutions to some specified combinatorial optimisation problem. In \cref{sec:QWOA} we describe the QWOA procedure and its quantum circuit implementation. In \cref{sec:walks}, we detail our method for quantum walking over a variety of combinatorial domains. Specifically, the domains applicable to this method are those with an associated \textit{indexing function}, which efficiently identifies each object with a unique integer index. Indexing combinatorial objects is called ranking in the literature, however we refer to it here as `indexing' to avoid confusion with ranking objects by their quality.  We show that the domain of combinatorial objects can be connected by any circulant graph, barring some minor restrictions, which would ensure a highly efficient quantum circuit implementation. We also show how to design a unitary that efficiently performs the indexing on computational basis states representing objects. 
	In \cref{sec:applications}, we give a number of applicable combinatorial domains along with their associated NP optimisation problems, including well-known problems such as the Travelling Salesman Problem and constrained portfolio optimisation. In \cref{sec:circuit}, we present specific quantum circuits to implement the indexing functions. Finally, we describe the relevance of the indexing unitary to quantum search in \cref{sec:search} and then make some concluding remarks.
	
	\section{Quantum walk-assisted optimisation algorithm\label{sec:QWOA}}
	
	\begin{figure*}
		\centering
		\[ \Qcircuit @C=1em @R=.5em {
			&&&&\push{\rule{2em}{0em}}&&\ustick{\hat{U}_Q(\gamma)}&&&&&&&&&\ustick{\hat{U}_W(t)}&&&&\push{\rule{2em}{0em}} \\
			&&&\lstick{\ket{\psi}} & \ctrl{1} & \qw & \qw & \qw & \ctrl{1} & \qw & \qw & \gate{\hat{U}_{\#}} & \gate{\hat{\mathcal{F}}} & \ctrl{1} & \qw & \qw & \qw & \ctrl{1} & \gate{\hat{\mathcal{F}}^\dag} & \gate{\hat{U}_{\#}^\dag} & \qw & {\ket{\psi'}}\\
			&&&\lstick{\ket{0}} & \multigate{3}{\hat{q}} & \qw & \gate{e^{-i \gamma \ket{1} \bra{1}}} & \qw & \multigate{3}{\hat{q}} & \qw & \qw & \qw & \qw & \multigate{3}{\hat{\lambda}} & \qw & \gate{e^{-i t \ket{1} \bra{1}}} & \qw & \multigate{3}{\hat{\lambda}} & \qw & \qw & \qw & {\ket{0}}\\
			\lstick{\rotatebox{90}{$\mathcal{O}(\text{polylog\,} N)$}} &&&\lstick{\ket{0}} & \ghost{\hat{q}} & \qw & \gate{e^{-2i \gamma \ket{1} \bra{1}}} & \qw & \ghost{\hat{q}} & \qw & \qw & \qw & \qw & \ghost{\hat{\lambda}} & \qw & \gate{e^{-2i t \ket{1} \bra{1}}} & \qw & \ghost{\hat{\lambda}} & \qw & \qw & \qw & {\ket{0}}\\
			&&&\lstick{\cdots} & & & \cdots & & & & & \cdots &  & & & \cdots & & & \cdots & & & \cdots \\
			&&&\lstick{\ket{0}} & \ghost{\hat{q}} & \qw & \gate{e^{-2^{k-1} i \gamma \ket{1} \bra{1}}} & \qw & \ghost{\hat{c}} & \qw & \qw & \qw & \qw & \ghost{\hat{\lambda}} & \qw & \gate{e^{-2^{k-1} i t \ket{1} \bra{1}}} & \qw & \ghost{\hat{\lambda}} & \qw & \qw & \qw & {\ket{0}} \\
			&&&&&&&&&&&&&&&&&&&& \\
			&&&&&&&&&&&&&&&\dstick{\exp(i \hat{C} t)}
			\gategroup{1}{5}{7}{10}{.7em}{--}\gategroup{1}{11}{7}{20}{.7em}{--}\gategroup{1}{13}{7}{19}{1.2em}{_\}}\gategroup{3}{2}{6}{2}{.7em}{\{}
		}\]
		\caption{An illustrative component of a QWOA circuit for optimisation over combinatorial domains having efficient indexing and un-indexing functions. The circuit performs $\ket{\psi'} = \hat{U}_W(t) \hat{U}_Q(\gamma) \ket{\psi}$ with $\mathcal{O}(\text{polylog} \, N)$ depth. Here the operator $\hat{q}$ evaluates the quality of a solution to $k$ bits of precision, and $\hat{\lambda}$ computes the eigenvalue of the circulant matrix to the same precision. The rotation angles $\gamma$ and $t$ can be varied by a classical optimiser.}
		\label{fig:qaoa_circuit}
	\end{figure*}
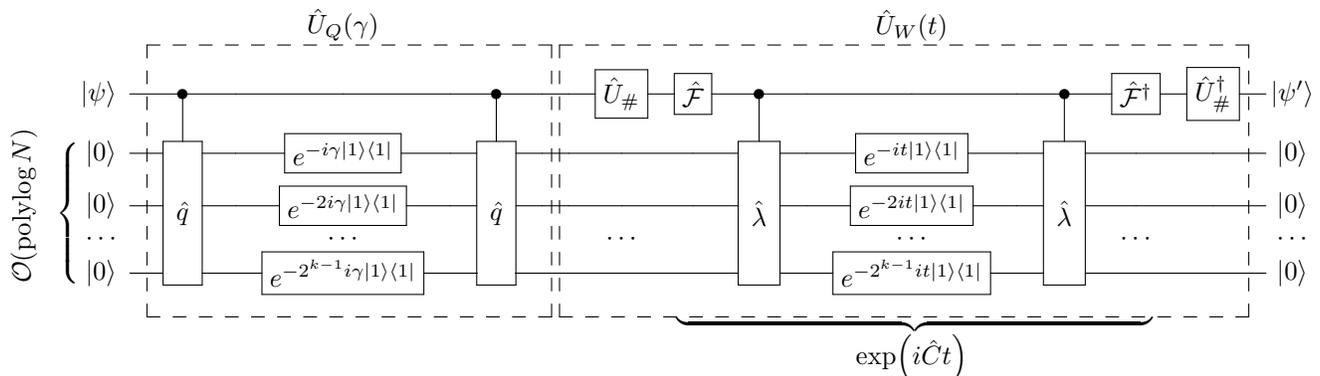
	
	QWOA uses the continuous time quantum walk as an ansatz, distinguished from the original derivation of the QAOA as a discretised adiabatic evolution. Specifically, in the QWOA framework the quantum system evolves as~\cite{Marsh2019}
	\begin{equation}
	\ket{\vec{\gamma}, \vec{t}} = \hat{U}_W(t_p) \hat{U}_Q(\gamma_p) \ldots \hat{U}_W(t_1) \hat{U}_Q(\gamma_1) \ket{s} \,.
	\end{equation}
	In a combinatorial optimisation context, the QWOA procedure can be interpreted as follows:
	\begin{enumerate}
		\item $\ket{s}$ is the initial state, which can be taken to be an equal superposition over all of the feasible combinatorial solutions.
		\item $\hat{U}_Q(\gamma)=e^{i \gamma \hat{Q}}$, where $\hat{Q}$ is a diagonal matrix with diagonal elements $q(x)$ corresponding to solution qualities with respect to the optimisation problem. As such it applies a phase shift to each combinatorial object, proportional to $\gamma$ and its \textit{quality}.
		\item $\hat{U}_W(t)$ performs a continuous-time quantum \textit{walk} for time $t$ over the combinatorial domain; its details will be discussed in the next section.
		\item The set of $2p$ parameters $\vec{\gamma} = (\gamma_1, \ldots, \gamma_p)$ and $\vec{t} = (t_1, \ldots, t_p)$ are chosen to maximise the expectation value $\bra{\vec{\gamma}, \vec{t}} \hat{Q} \ket{\vec{\gamma}, \vec{t}}$, representing the average measured solution quality. 
	\end{enumerate}
	The QWOA state evolution consists of an interleaved series of phase shifts, which introduce a bias to solutions dependent on their quality, and quantum walks to mix amplitude between solutions. A higher choice of $p$ leads to better solutions at the cost of a longer quantum computation with more variational parameters. In the general case, a classical optimiser can be used to vary the parameters $\vec{\gamma}$ and $\vec{t}$ such that the expectation value of the solution quality is maximised. A circuit diagram for a component of the QWOA is illustrated in \cref{fig:qaoa_circuit}, consisting of one solution quality dependent phase shift $\hat{U}_Q(\gamma)$ followed by a quantum walk over the valid solutions $\hat{U}_W(t)$.
	
	The $\hat{U}_Q(\gamma)$ component of the QWOA circuit is straightforward to implement for any combinatorial optimisation problem in the NPO complexity class. Put another way, given a combinatorial object $x$, we require that the solution quality $q(x)$ can be efficiently computed. If so, there is an efficient quantum circuit that implements the desired solution quality-dependent phase shift \cite{childs2004quantum}. We show this implementation in the left dashed box of \cref{fig:qaoa_circuit}. With this in mind, the remainder of this paper proposes a generic approach to efficiently implementing $\hat{U}_W(t)$ over a wide range of combinatorial domains. 
	
	
	\section{Quantum walks over combinatorial domains\label{sec:walks}}
	
\begin{figure}[b]
		\subfloat[]{
			\centering
			\includegraphics[width=0.5\linewidth]{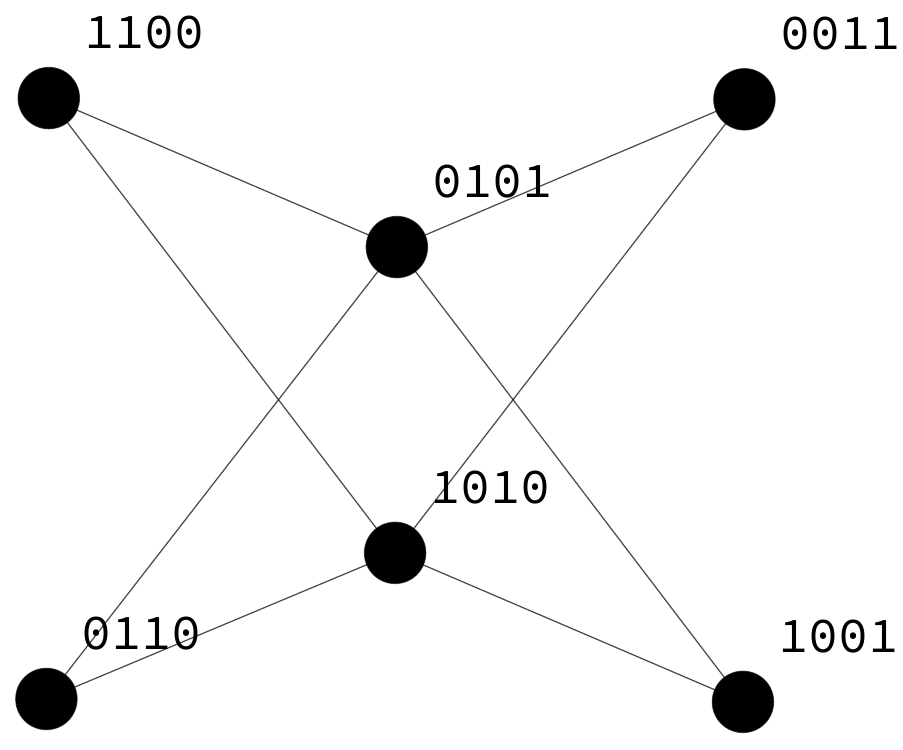}
		}
		\subfloat[]{
			\includegraphics[width=0.5\linewidth]{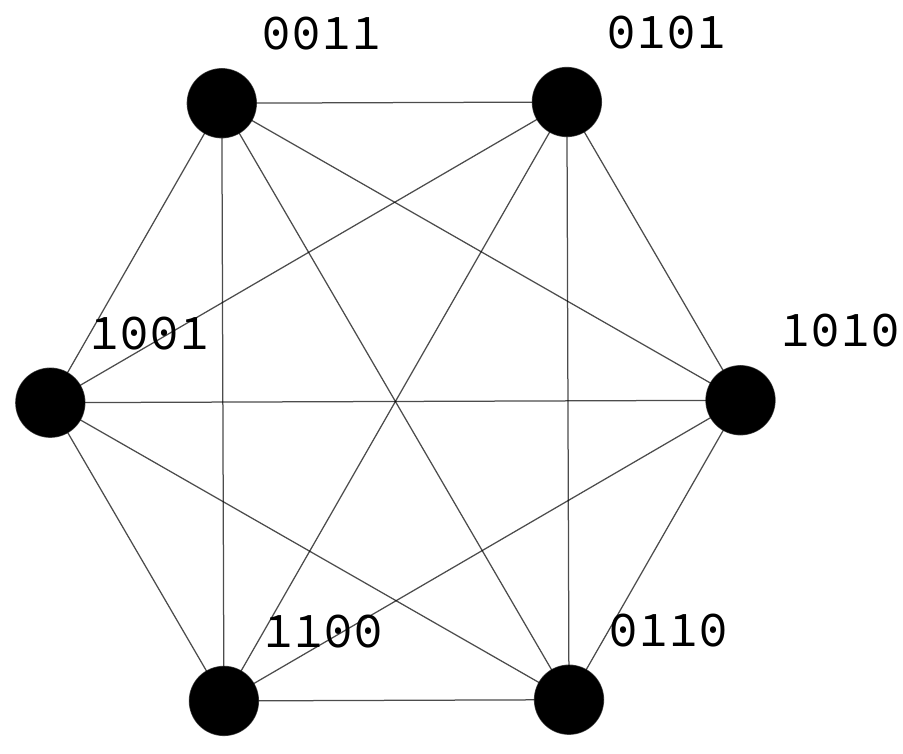}
        }
		\caption{(a) $XY$ model connectivity of length-$4$ bitstrings with exactly $2$ bits on. The two central solutions have double the degree of the outer ones, introducing bias. (b) Using our proposed approach with the complete graph maintains symmetry amongst solutions.}
		\label{fig:xycompletecompare}
	\end{figure}

	Consider a set of $M$ combinatorial objects, each encoded as an $n$-bit string. An example is $k$-combinations of a set $S = \{s_0, s_1, \ldots, s_{n-1} \}$, where each combinatorial object corresponds to a unique $k$-combination of $S$. A straightforward encoding of a specific $k$-combination as a length-$n$ binary string is $x = x_0 x_1 \ldots x_{n-1}$, where $x_j = 1$ iff $s_j$ is selected as part of the combination. In this case, the aim is to perform a quantum walk over a graph connecting each of the $M = {n \choose k}$ possible $k$-combinations of $S$. Current approaches resort to sparse or approximate Hamiltonian simulation of the $XY$ model Hamiltonian \cite{Hadfield2019,XYMixers2019}, which is not ideal as its corresponding quantum circuit changes with the value of the walk time $t$ (requiring a longer circuit to approximate the quantum walk dynamics for a longer time), and furthermore the the $XY$ model does not maintain symmetry amongst the relevant vertices, introducing bias into the optimisation algorithm. This is illustrated in \cref{fig:xycompletecompare}. Our method provides the ability to resolve this issue with (for example) the complete graph or the cycle graph, preserving symmetry.
	
	We choose to focus on circulant connectivity because all circulant graphs are diagonalised by the Fourier transform. Thus, the highly efficient Quantum Fourier Transform (QFT) can be used to diagonalise any choice of circulant graph, whilst the circuit structure remains largely unchanged \cite{Qiang2016,Zhou2017,Zhou2017qft,Qiang2018}. The first step of our approach is to unitarily map the $M$ binary strings corresponding to combinatorial objects, which are `scattered' in a larger space of $2^n$ binary strings, to a canonical subspace of the first $M$ binary strings (i.e. $0$ through to $M-1$). This is done using a so-called `indexing unitary' $U_{\#}$. We then describe how to perform a continuous time quantum walk over the first $M$ computational basis states connected as a circulant graph, where $M$ is not necessarily a power of 2. After `un-indexing' using $U_{\#}^\dag$, a CTQW over the circulantly connected combinatorial objects has been performed. Our approach is summarised with
	\begin{equation}
	\hat{U}_W(t) = \hat{U}_{\#}^\dag \hat{\mathcal{F}}_M \exp(i \hat{\lambda} t) \hat{\mathcal{F}}^\dag_M \hat{U}_{\#} \, ,
	\end{equation}
	where $\hat{U}_{\#}$ is the indexing unitary, $\hat{\mathcal{F}}_M$ is the Quantum Fourier Transform modulo $M$ \cite{Kitaev1995}, and $\hat{\lambda}$ is a diagonal matrix containing the $M$ eigenvalues of the circulant adjacency matrix, which describes the connectivity of the objects. In the following subsections we describe each component of the overall quantum walk, and the corresponding circuit implementation.
	
	\subsection{\label{sec:indexing}Indexing with $\hat{U}_{\#}$}
	
	
	Let $S = \{s_1, s_2, \ldots, s_M\}$ be a set of bitstrings encoding combinatorial objects. Assume the combinatorial objects can be encoded using $n$ qubits. For example, $k$-combinations from $[n]=\{0,1,\ldots,n-1\}$ may be represented by an $n$-bit string with exactly $k$ bits set. We wish to perform a continuous time quantum walk (CTQW) over $S$. Consider a bijective function $id : S \rightarrow [M]$ that identifies each element of $S$ with its unique integer `index', where both $id$ and $id^{-1}$ are efficiently computable. Typical indexing functions usually order lexicographically, although for our purposes the manner of ordering is not relevant. Then with application of a unitary that performs indexing on the subspace of valid combinatorial bitstrings, quantum walking over $S$ is reduced to quantum walking over $[M]$.
	
\begin{figure}[h!]
		\centering
		\[ \Qcircuit @C=1em @R=.5em {
			\lstick{\ket{x}} & \multigate{1}{\hat{U}_{id}} & \qswap & \multigate{1}{\hat{U}_{id^{-1}}} & \qw & \rstick{\ket{id(x)}} \\
			\lstick{\ket{0}} & \ghost{\hat{U}_{id}} & \qswap \qwx & \ghost{\hat{U}_{id^{-1}}} & \qw & \rstick{\ket{0}}
		}
		\]
		\caption{Circuit for $\hat{U}_{\#}$, using an ancilla register of the same size as the input register.}
		\label{fig:indexcircuit}
	\end{figure}
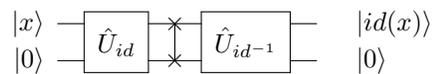

	We now briefly explain how to construct the indexing unitary $\hat{U}_{\#}$, given classical algorithms for indexing and un-indexing. Since we assume we have an efficient classical algorithm for indexing, we can construct a reversible circuit to perform $\hat{U}_{id} \ket{x} \ket{z} = \ket{x} \ket{z \oplus id(x)}$. Similarly, we can construct a unitary for un-indexing, $\hat{U}_{id^{-1}} \ket{id(x)} \ket{z} = \ket{id(x)} \ket{z \oplus x}$. It is straightforward to check that
	\begin{equation}
	\hat{U}_{\#} \ket{x} \ket{0} \equiv U_{id^{-1}} \hat{\mathcal{S}} \hat{U}_{id} \ket{x} \ket{0} = \ket{id(x)} \ket{0} \, ,
	\end{equation}
	where $\hat{\mathcal{S}}$ swaps the first and second registers. We give the circuit diagram in \cref{fig:indexcircuit}. Clearly, un-indexing is performed with $\hat{U}_{\#}^\dag$.
	
	The indexing algorithm depends on the specific combinatorial family. However, there is a general framework by \citet{WILF1977281} that applies when the combinatorial objects are constructed recursively. For example an arbitrary $n$-choose-$k$ combination is built from the $\binom{n-1}{k}$ subsets not including $n$, and the $\binom{n-1}{k-1}$ subsets that do include $n$. In this sense, the construction of a combinatorial object can be thought of as a certain path on a directed graph, with each step partially building the object. The initial point of the walk -- for example $(n, k)$ in the lattice describing all set combinations -- describes the so-called \textit{order} of the object. In this framework, the indexing algorithm has complexity scaling as a polynomial in the object's order. As per \cite{WILF1977281}, some illustrative combinatorial families for which this framework applies include $k$-subsets of $[n]$, permutations of $[n]$ with $k$ cycles, and $k$-partitions of $[n]$. Associated with each of these example families is a formula to obtain the number of objects of a given order -- the binomial coefficients, the Stirling numbers of the first kind, and the Bell numbers respectively. This is a property induced by \citeauthor{WILF1977281}'s framework, and applies to all the combinatorial families within it. Thus for our purposes, the number of objects $M$ is always explicitly known (or can be determined efficiently).
	
	Finally, given the ability to index objects of a given order, the indexing algorithm can be extended to handle a range of orders $a \leq k \leq b$. This makes our framework naturally applicable to many more relevant combinatorial domains. For example, to index a combination $\vec{c}=(c_0, \ldots, c_{j-1})$ of $[n]$ with respect to all combinations of $\leq K$ elements,
	\begin{equation}
		id_{\leq K}(\vec{c}) = id_{\text{len}(\vec{c})}(\vec{c}) + \sum\limits_{k=0}^{\text{len}(\vec{c})-1} \binom{n}{k}
	\end{equation}
	where $\text{len}(\vec{c})$ is the order of $\vec{c}$, i.e. the number of elements in the combination. 
	
	\subsection{Preparation of the initial state}
	
	In the original QAOA, one prepares an equal superposition over each of the $N=2^n$ computational basis states. However, in general, up to half of these bit-strings do not correspond to a valid combinatorial object contained in the set $S$, i.e. $M<N$.
	
	Instead, to prepare the equal superposition $\ket{s}$ over the valid combinatorial objects, we first apply the Quantum Fourier Transform modulo $M$ to the $\ket{0}^{\otimes n}$ state. There are highly depth-efficient quantum circuits that can be used to implement the Fourier transform with arbitrary modulus \cite{Cleve2000}. This creates the superposition $(1/\sqrt{M}) \sum_{x=0}^{M-1} \ket{x}$. Then the un-indexing unitary $U_{\#}^{-1}$ can be carried out, thus efficiently preparing the desired initial state
	\begin{equation}
	\ket{s} = \frac{1}{\sqrt{M}} \sum\limits_{x \in S} \ket{x} \, .
	\end{equation}
	
	\subsection{Quantum walk with $\exp(i \hat{C} t)$}
	
	We now connect the indices in a circulant manner, defining a circulant matrix $\hat{C}$ of size $M \times M$. This circulant matrix is diagonalised by the Fourier matrix $\hat{\mathcal{F}}_M$. Therefore the quantum walk over the graph with adjacency matrix $C$ is
	\begin{equation}
	    \exp(i \hat{C} t) = \hat{\mathcal{F}}_M \exp(i \hat{\lambda} t) \hat{\mathcal{F}}^\dag_M \, .
	\end{equation}
	Again, the Quantum Fourier Transform modulo $M$ can be used to efficiently perform the diagonalisation.
	
	The unitary $\exp(i \hat{\lambda} t)$ can be implemented efficiently using well-known methods from \cite{childs2004quantum}, similar to $\hat{U}_Q(\gamma)$. We require that the diagonal elements of $\hat{\lambda}$ be efficiently computable, which is the case for graphs having efficiently computable eigenvalues. This holds for circulant graphs when the maximum degree grows polynomially, or when the eigenvalues are known in closed-form (e.g. complete, cycle and M\"obius ladder graphs). Since each unitary can be implemented efficiently, the entire walk is efficient. It is anticipated that a low choice of $p$, leading to a relatively shallow circuit, will have a quantum advantage for near-term NISQ applications as discussed in \cite{Bravyi308}. 
	
	\section{Applications\label{sec:applications}}
	
	In the following, we give an overview of a number of combinatorial structures and give their associated indexing functions. A continuous time quantum walk can be efficiently performed over each of following domains using the above-described scheme.
	
	There are a wide range of integer sequences with indexing and un-indexing algorithms  (or equivalently an efficient closed-form expression for the $n$th element of the sequence $a_n$, where the inverse operation is also efficiently computable). A comprehensive list of such sequences can be found on The On-Line Encyclopedia of Integer Sequences (OEIS) \cite{oeis}, with some examples given below. For brevity we do not give details on the corresponding un-indexing functions; their implementations are similar to indexing with the same runtime complexity, and are available in the references provided in the below sections.
	
	\subsection{Set $k$-combinations}
	
	Let a binary string $x=x_1 \ldots x_n$ denote a combination, where $x_j = 1$ iff the $j$th element is selected. Then an efficient indexing algorithm, to index a given $k$-combination amongst all other $k$-combinations, is given in \cref{alg:combinationindex}.
	\begin{algorithm}[H]
		\caption{$\text{INDEX\_COMB}(x, k)$}
		\label{alg:combinationindex}
		\begin{algorithmic}
			\State $c \gets $ list of $j$ where $x_j = 1$ 
			\State sort $c$ in ascending order
			\State \Return sum from $j=1$ to $k$ of ${c[j] \choose j}$
		\end{algorithmic}
	\end{algorithm}
	
\begin{figure*}[t]
		\subfloat[]{
			\centering
		    \includegraphics[width=0.3\linewidth]{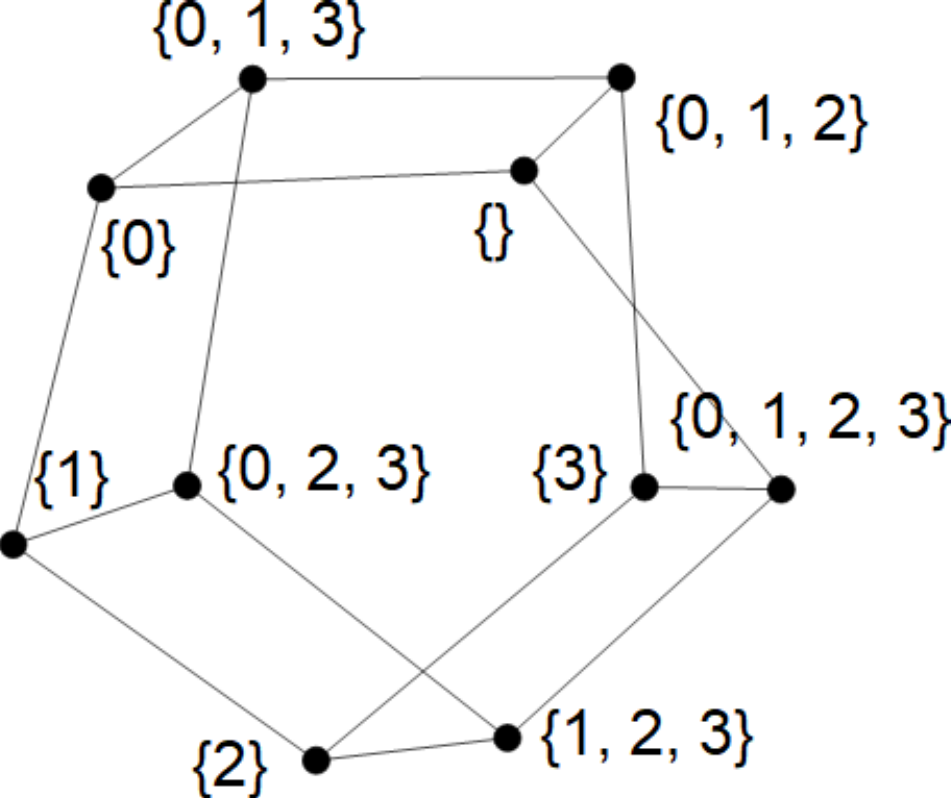}
		}
		\subfloat[]{
		    \raisebox{\depth}{
		    \centering
		    \includegraphics[width=0.3\linewidth,valign=m]{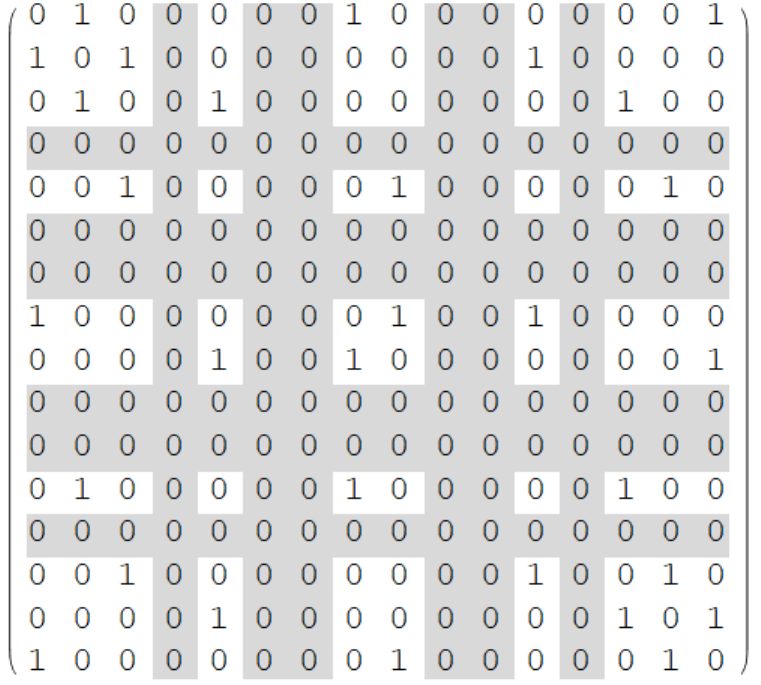}
		    $\longrightarrow$
		    \includegraphics[width=0.3\linewidth,valign=m]{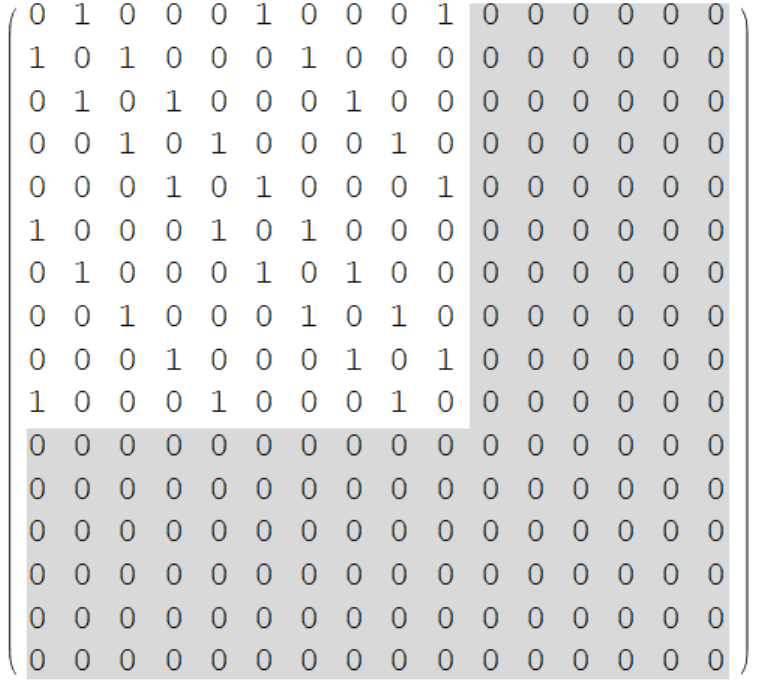}
		    }
		}
		\caption{(a) A M\"obius ladder graph connecting subsets of $\{0,1,2,3\}$ that do not contain exactly $2$ elements. (b) The adjacency matrix before and after applying the indexing unitary. Rows coloured grey correspond to invalid states.}
		\label{fig:equal_hamming_weight_graph}
	\end{figure*}

	As discussed in \cref{sec:indexing}, the $k$-combination indexing function can be `wrapped' to index combinations amongst others of chosen sizes. As a non-trivial example, in \cref{fig:equal_hamming_weight_graph} we show a M\"obius ladder graph connecting all subsets of $\{0,1,2,3\}$ \textit{except} those with exactly two elements.
	
	By choosing $n$ to be even and $k=n/2$, this is the domain of the graph partitioning problem. Similarly, this indexing algorithm can be applied to the critical node detection problem \cite{LALOU201892}. Another real-world application is to constrained financial portfolio optimisation in finance, where up to $k$ assets are selected to create a financial portfolio with minimum risk and/or maximum return.
	
	In \cref{sec:circuit}, we give two quantum circuits to perform efficient $k$-combination ranking, depending on how the combinations are represented in the quantum register.
	
	
	
	
	\subsection{Permutations}
	
	A permutation can be indexed in linear time using \cref{alg:permutationindex} \cite{MYRVOLD2001281}, where $\pi$ is a permutation of $[n]$ and $\pi^{-1}$ is the inverse permutation (which can be computed in linear time). 
	
	\begin{algorithm}[H]
		\caption{$\text{INDEX\_PERM}(n, \pi, \pi^{-1})$}
		\label{alg:permutationindex}
		\begin{algorithmic}
			\If{$n = 1$} \State \Return 0 \EndIf
			\State $s \gets \pi[n-1]$
			\State swap $\pi[n-1]$ and $\pi[\pi^{-1}[n-1]]$
			\State swap $\pi^{-1}[s]$ and $\pi^{-1}[n-1]$
			\State \Return $s + n \times \text{INDEX\_PERM}(n-1, \pi, \pi^{-1})$
		\end{algorithmic}
	\end{algorithm}
	
	As an example, the well-known NP-hard Travelling Salesman Problem has domain corresponding to the $k!$ possible permutations of visiting each of $k$ cities exactly once and then returning to the starting city. A straightforward encoding is to use $k \ceil{\log k}$ qubits, where each block of $\ceil{\log k}$ qubits represents the next city to visit. The indexing unitary $U_{\#}$ can then be used to map this encoding of a tour to an integer from $0$ to $(k!-1)$, and vice versa.
	
	In \cref{sec:circuit}, we provide a simple quantum circuit to perform efficient permutation ranking.

	\subsection{Lattice paths}
	
	A Dyck path is a lattice path from $(0, 0)$ to $(n, n)$ which never goes above the diagonal $y=x$. They are an example of a combinatorial structure characterised by the Catalan numbers, a sequence appearing in a number of combinatorial applications. Catalan numbers have an efficient indexing function provided in \cref{alg:catalanindex}, where the function $\text{NUM\_DYCK}(i, j)$ counts the total number of Dyck paths between $(0, 0)$ and $(i, j)$ \cite{Kasa2010}. 
	\begin{algorithm}[H]
		\caption{$\text{INDEX\_CATALAN}(x)$}
		\label{alg:catalanindex}
		\begin{algorithmic}
			\State $n \gets $ number of bits of $x$
			\State $c_1 \gets 2$
			\For{$j = 2$ to $n$}
			\State $c_j \gets \max(x_{j-1} + 1, 2j)$
			\EndFor
			\State $r \gets 1$
			\For{$j = 1$ to $n - 1$}
			\For{$k = c_j$ to $b_j - 1$}
			\State $r \gets r + \text{NUM\_DYCK}(n-j,n+j-k)$
			\EndFor
			\EndFor
			\State \Return $r$
		\end{algorithmic}
	\end{algorithm}
	
	There are other lattice path applications. For example, an arbitrary length-$n$ path in 3D space can be characterised by an $n$-letter word where each letter is chosen from an alphabet of six directions. Any `word' composed of a sequence of letters from some specified alphabet can be indexed, and constraints on some/all of the letters can also be incorporated \cite{Loehr2011}.
	
	\section{Quantum circuits for indexing\label{sec:circuit}}
	
\begin{figure*}[t!]
	    \centering
        \subfloat[]{
            \begin{minipage}[c]{.24\textwidth}
                $\Qcircuit @C=.5em @R=.5em {
        	        \lstick{\ket{x}} & {/} \qw & \multigate{1}{\binom{\cdot}{m}} & \qw & \rstick{\ket{x}} \\
        	        \lstick{\ket{0}} & {/} \qw & \ghost{\binom{\cdot}{m}} & \qw & \rstick{\ket{\binom{x}{m}}}
        	    }$
            \end{minipage}
    	}
    	\subfloat[]{
    	    \begin{minipage}[c]{.24\textwidth}
                $ \Qcircuit @C=.5em @R=.5em {
        	        \lstick{\ket{x}} & {/} \qw & \multigate{1}{\binom{m}{\cdot}} & \qw & \rstick{\ket{x}} \\
        	        \lstick{\ket{0}} & {/} \qw & \ghost{\binom{m}{\cdot}} & \qw & \rstick{\ket{\binom{m}{x}}}
        	    }$
    	   \end{minipage}

    	}
    	\subfloat[]{
    	    \begin{minipage}[c]{.24\textwidth}
                $\Qcircuit @C=.5em @R=.5em {
        	        \lstick{\ket{x}} & {/} \qw & \multigate{1}{+} & \qw & \rstick{\ket{x}} \\
        	        \lstick{\ket{y}} & {/} \qw & \ghost{+} & \qw & \rstick{\ket{x+y}}
        	    }$
    	    \end{minipage}
    	}
    	\subfloat[]{
    	    \begin{minipage}[c]{.24\textwidth}
                $\Qcircuit @C=.5em @R=.5em {
        	        \lstick{\ket{x}} & {/} \qw & \ctrl{1} & \qw & \rstick{\ket{x}} \\
        	        \lstick{\ket{y}} & {/} \qw & \ctrl{1} & \qw & \rstick{\ket{y}} \\
        	        \lstick{\ket{z}} & \qw & \gate{>} & \qw & \rstick{\ket{z \oplus (x>y)}}
        	    }$
            \end{minipage}
    	}
    	\caption{Notation for the unitary operations required to implement $\hat{U}_{id}$ for $n$-choose-$k$-combinations and $n$-permutations. Fundamental gate counts: (a) $\mathcal{O}(k \log n)$ (b) $\mathcal{O}(n \log n)$ (c) $\mathcal{O}(\log n)$ (d) $\mathcal{O}(\log n)$.}
    	\label{fig:notation}
	\end{figure*}
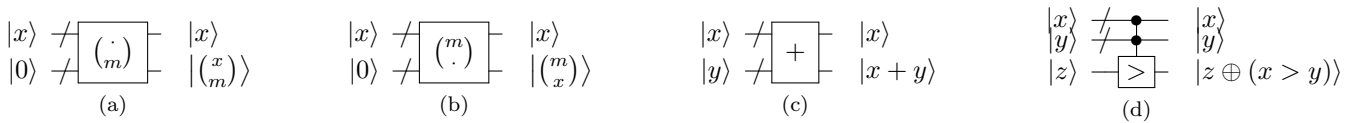

	In this section, we give some quantum circuits for $U_{id}$. We use the notation shown in \cref{fig:notation} for the required unitary operations.
	
	\subsection{Combinations}
	
	Here we present quantum circuits for indexing the two main representations of $k$-combinations. Consider a selection of $k$ integers from the set $[n]$. The first representation is to use a binary string $x$ of length $n$, where $x_i = 1$ if $i$ is selected. This clearly requires $n$ bits for any $k$. The second representation is to directly represent the $k$ combination as a sequence of $k$ integers, requiring $O(k \log n)$ space. Both circuits follow the indexing method provided in \cref{alg:combinationindex}. To un-index, a circuit to prepare uniform superpositions of Dicke states can be used, such as \cite{Bartschi2019}.

	The first quantum circuit for indexing $k$-combinations of $[n]$ is given in \cref{fig:combinationcircuit}. It consists of an input register $\ket{x}$ of $n$ qubits and an output register of the same size, as well as two ancilla registers of $\mathcal{O}(\log k)$ and $n$ qubits respectively. The fundamental gate count is $\mathcal{O}(n k \log n)$.
	
\begin{figure*}[t]
		\centering
		\[ \Qcircuit @C=.5em @R=.8em {
			\lstick{\ket{c_0}} & {/} \qw & \qswap &\qw & \qw & \qw & \qswap & \qw & \qw & \qw & \qw & \qw & \qw & \push{\rule{1em}{0em}} & \qw & \qw & \qw & \qw & \rstick{\ket{c_0}} \\
			\lstick{\ket{c_1}} & {/} \qw & \qw & \qw & \qw & \qw & \qw & \qswap & \qw & \qw & \qw &\qswap & \qw & \push{\rule{1em}{0em}} & \qw & \qw & \qw & \qw & \rstick{\ket{c_1}}\\
			\lstick{\cdots} & \push{\rule{0em}{1em}} & & & & & & & & & & & & \cdots \\
			\lstick{\ket{c_{k-1}}} & {/} \qw & \qswap\qwx[-3] & \multigate{1}{\binom{\cdot}{1}} & \qw & \multigate{1}{\binom{\cdot}{1}} & \qswap\qwx[-3]& \qswap\qwx[-2] &  \multigate{1}{\binom{\cdot}{2}} & \qw & \multigate{1}{\binom{\cdot}{2}} &\qswap\qwx[-2] & \qw & \push{\rule{1em}{0em}} & \multigate{1}{\binom{\cdot}{k}} & \qw & \multigate{1}{\binom{\cdot}{k}} & \qw & \rstick{\ket{c_{k-1}}}\\
			\lstick{\ket{0}} & {/} \qw & \qw & \ghost{\binom{\cdot}{1}} & \multigate{1}{+} & \ghost{\binom{\cdot}{1}} & \qw & \qw & \ghost{\binom{\cdot}{2}} & \multigate{1}{+} & \ghost{\binom{\cdot}{2}} & \qw & \qw & \push{\rule{1em}{0em}} & \ghost{\binom{\cdot}{k}} & \multigate{1}{+} & \ghost{\binom{\cdot}{k}} & \qw & \rstick{\ket{0}}\\
			\lstick{\ket{0}} & {/} \qw & \qw & \qw & \ghost{+} & \qw & \qw & \qw & \qw & \ghost{+} & \qw & \qw & \qw & \push{\rule{1em}{0em}} & \qw & \ghost{+} & \qw & \qw & \rstick{\ket{id(c)}}\\
		}\]
		\caption{Quantum circuit for indexing a $k$-combination represented by $(c_0, \ldots, c_{k-1})$ with elements in ascending order.}
		\label{fig:combinationcircuit2}
	\end{figure*}
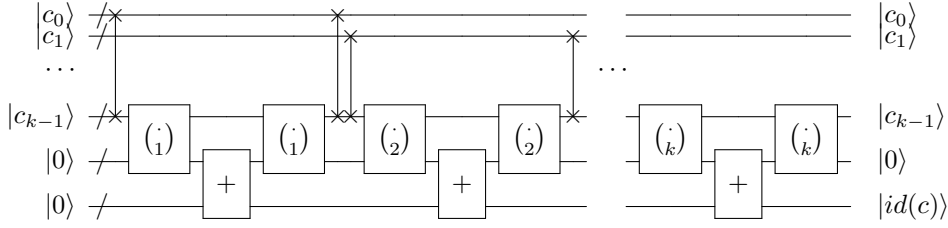

	The quantum circuit for the second representation is shown in \cref{fig:combinationcircuit2}. The input, output and ancilla register are of size $\mathcal{O}(k \log n)$. The fundamental gate count is $\mathcal{O}(k^2 \log n)$.
	
	It depends upon the application as to which representation is more appropriate. If $k$ scales proportionally to $n$, then the binary representation is more space and time-efficient. Otherwise, if $k$ is small (or constant) then the second quantum circuit is appropriate.
	
	\begin{figure*}[t]
		\centering
		\[ \Qcircuit @C=.5em @R=.5em {
			\lstick{\ket{x_0}} & \qw & \qswap & \qw & \qswap & \ctrl{4} & \qw & \ctrl{4} & \qw & \qw & \qw & \qw & \qw & \qw & \qw & \push{\rule{1em}{0em}} & \qw & \qw & \qw & \qw & \rstick{\ket{x_0}} \\
			\lstick{\ket{x_1}} & \qw & \qw & \qw & \qw & \qw & \qw & \qw & \qswap & \qw & \qswap & \ctrl{3} & \qw & \ctrl{3} & \qw & \push{\rule{1em}{0em}} & \qw & \qw & \qw & \qw & \rstick{\ket{x_1}} \\
			\lstick{\cdots} && \push{\rule{0em}{1em}} & & & & & & & & & & & & & \cdots \\
			\lstick{\ket{x_{n-1}}} & \qw & \qswap\qwx[-3] & \multigate{1}{+} & \qswap\qwx[-3] & \qw & \qw & \qw & \qswap\qwx[-2] & \multigate{1}{+} & \qswap\qwx[-2] & \qw & \qw & \qw & \qw & \push{\rule{1em}{0em}} & \ctrl{1} & \qw & \ctrl{1} & \qw & \rstick{\ket{x_{n-1}}}\\
			\lstick{\ket{0}} & {/} \qw & \qw & \ghost{+} & \qw & \multigate{1}{\binom{0}{\cdot}} & \qw & \multigate{1}{\binom{0}{\cdot}} & \qw & \ghost{+} & \qw & \multigate{1}{\binom{1}{\cdot}} & \qw & \multigate{1}{\binom{1}{\cdot}} & \qw & \push{\rule{1em}{0em}} & \multigate{1}{\binom{n-1}{\cdot}} & \qw & \multigate{1}{\binom{n-1}{\cdot}} & \qw & \rstick{\ket{k}} \\
			\lstick{\ket{0}} & {/}\qw & \qw & \qw & \qw & \ghost{\binom{0}{\cdot}} & \multigate{1}{+} & \ghost{\binom{0}{\cdot}} & \qw & \qw & \qw & \ghost{\binom{1}{\cdot}} & \multigate{1}{+} & \ghost{\binom{1}{\cdot}} & \qw & \push{\rule{1em}{0em}} & \ghost{\binom{n-1}{\cdot}} & \multigate{1}{+} & \ghost{\binom{n-1}{\cdot}} & \qw & \rstick{\ket{0}} \\
			\lstick{\ket{0}} & {/} \qw &\qw & \qw & \qw & \qw & \ghost{+} & \qw & \qw & \qw & \qw & \qw & \ghost{+} & \qw & \qw & \push{\rule{1em}{0em}} & \qw & \ghost{+} & \qw & \qw & \rstick{\ket{id(x)}}
		}\]
		\caption{Quantum circuit for indexing a $k$-combination represented by $x=x_0 \ldots x_{n-1}$, where $x_i=1$ if $i$ is selected.}
		\label{fig:combinationcircuit}
	\end{figure*}
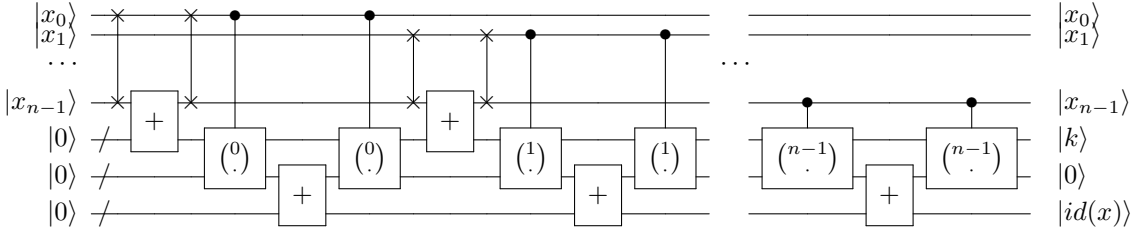
	
	\subsection{Permutations}
	
	To design a quantum circuit for indexing permutations, we use the expression
	\begin{equation}
	id(\pi) = (\ldots (d_0 (n-1) + d_1) \cdot (n-2) + \ldots + d_{n-2}) \cdot 1
	\end{equation}
	where $d_i$ is the $i$th digit of the Lehmer code corresponding to $\pi$. Explicitly, $d_i = |\{j>i : \pi_j < \pi_i \}|$. Note that to un-index permutations a circuit to prepare superpositions of permutations can be used, such as \cite{Chiew2019}.
	
	Consider a permutation on $[n]$, $\pi = \pi_0 \ldots \pi_{n-1}$. The permutation indexing circuit requires an input register $\ket{\pi} = \ket{\pi_0} \ldots \ket{\pi_{n-1}}$ of size $\mathcal{O}(n \log n)$, to hold each of the $n$ elements $\ket{\pi_i}$ having size $\mathcal{O}(\log n)$. An output register of the same size to hold $id(\pi)$, and an additional single qubit ancilla, are also required.
	
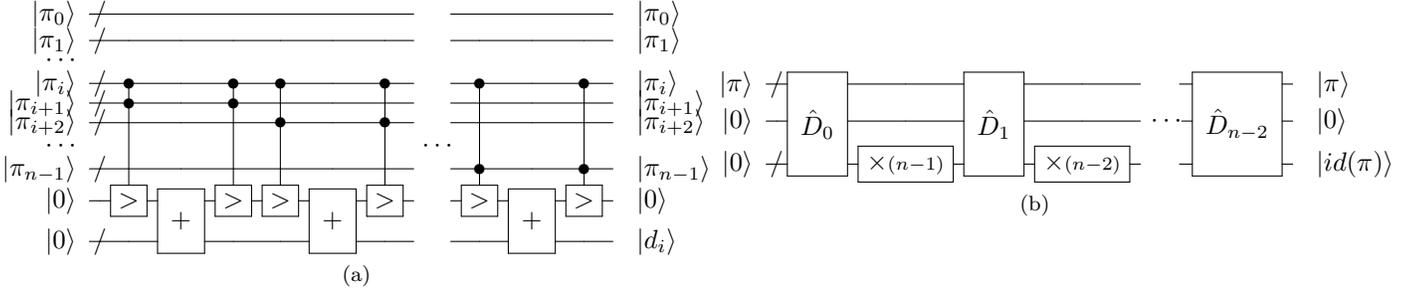
\begin{figure*}[t!]
		\subfloat[]{
		    \begin{minipage}[c]{.49\textwidth}
		    $\Qcircuit @C=.4em @R=.5em {
				\lstick{\ket{\pi_0}} & {/} \qw & \qw & \qw & \qw & \qw & \qw & \qw & \qw & \push{\rule{1em}{0em}} & \push{\rule{0em}{1em}}\qw & \qw & \qw & \qw & \rstick{\ket{\pi_0}} \\
				\lstick{\ket{\pi_1}} & {/} \qw & \qw & \qw & \qw & \qw & \qw & \qw & \qw & \push{\rule{1em}{0em}} & \qw & \qw & \qw & \qw & \rstick{\ket{\pi_1}} \\
				\lstick{\cdots} & \push{\rule{0em}{0.5em}} \\
				\lstick{\ket{\pi_i}} & {/} \qw & \ctrl{1} & \qw & \ctrl{1} & \ctrl{2} & \qw & \ctrl{2} & \qw & \push{\rule{1em}{0em}} & \ctrl{4} & \qw & \ctrl{4} & \qw & \rstick{\ket{\pi_i}} \\
				\lstick{\ket{\pi_{i+1}}} & {/} \qw & \ctrl{4} & \qw & \ctrl{4} & \qw & \qw & \qw & \qw & \push{\rule{1em}{0em}} & \qw & \qw & \qw & \qw & \rstick{\ket{\pi_{i+1}}} \\
				\lstick{\ket{\pi_{i+2}}} & {/} \qw & \qw & \qw & \qw & \ctrl{3} & \qw & \ctrl{3} & \qw & \push{\rule{1em}{0em}} & \qw & \qw & \qw & \qw & \rstick{\ket{\pi_{i+2}}}\\
				\lstick{\cdots} & \push{\rule{0em}{0.5em}} & & & & & & & & \cdots \\
				\lstick{\ket{\pi_{n-1}}} & {/} \qw & \qw & \qw & \qw & \qw & \qw & \qw & \qw & \push{\rule{1em}{0em}} & \ctrl{1} & \qw & \ctrl{1} & \qw & \rstick{\ket{\pi_{n-1}}} \\
				\lstick{\ket{0}} & \qw & \gate{>} & \multigate{1}{+} & \gate{>} & \gate{>} & \multigate{1}{+} & \gate{>} & \qw & \push{\rule{1em}{0em}} & \gate{>} & \multigate{1}{+} & \gate{>} & \qw & \rstick{\ket{0}} \\
				\lstick{\ket{0}} & {/} \qw & \qw & \ghost{+} & \qw & \qw & \ghost{+} & \qw & \qw & \push{\rule{1em}{0em}} & \qw & \ghost{+} & \qw & \qw & \rstick{\ket{d_i}}
			}$
			\end{minipage}
			\label{fig:permutationsubcircuit}
		}
		\subfloat[]{
			\begin{minipage}[c]{.49\textwidth}
			$\Qcircuit @C=.4em @R=.5em {
				\lstick{\ket{\pi}} & {/} \qw & \multigate{2}{\hat{D}_0} & \qw & \multigate{2}{\hat{D}_1} & \qw & \qw & \push{\rule{1em}{0em}} & \multigate{2}{\hat{D}_{n-2}} & \qw & \rstick{\ket{\pi}}\\
				\lstick{\ket{0}} & \qw & \ghost{\hat{D}_0} & \qw & \ghost{\hat{D}_1} & \qw & \qw & \push{\cdots}\ & \ghost{\hat{D}_{n-2}} & \qw & \rstick{\ket{0}} \\
				\lstick{\ket{0}} & {/} \qw & \ghost{\hat{D}_0} & \gate{\parbox{3em}{\centering $\times\scriptstyle(n-1)$}} & \ghost{\hat{D}_1} & \gate{\parbox{3em}{\centering $\times\scriptstyle(n-2)$}} & \qw & \push{\rule{1em}{0em}} & \ghost{\hat{D}_{n-2}} & \qw & \rstick{\ket{id(\pi)}}
			}$
			\end{minipage}
		    \label{fig:permutationcircuit}
    	}
    	\caption{(a) The quantum circuit for unitary operator $\hat{D}_i$, where $\hat{D}_i \ket{\pi} \ket{0} = \ket{\pi} \ket{d_i}$. Here, $d_i$ is the $i$th digit of the Lehmer code corresponding to permutation $\pi$. (b) The quantum circuit for indexing a given permutation $\pi$.}
	\end{figure*}

	We first give a quantum circuit to implement a Lehmer operator $D_i$ in \cref{fig:permutationsubcircuit}. This circuit performs $\ket{\pi} \ket{0} \mapsto \ket{\pi} \ket{d_i}$, using $\mathcal{O}(n \log n)$ gates.
	
	Using the Lehmer sub-circuit, indexing becomes straightforward as per \cref{fig:permutationcircuit}. There are $\mathcal{O}(n)$ uses of $\hat{D}_i$, leading to a gate count of $\mathcal{O}(n^2 \log n)$ to index permutations in this way. Although there is a classically linear approach to indexing given in \cref{alg:permutationindex}, it appears to become quadratic when quantised due to requiring access to the input register in a superposition of locations, leading to an additional linear overhead \cite{Berry2018}.
	
	\section{Quantum search\label{sec:search}}
	
	Finally, we briefly make explicit the connection with quantum search. The aim here is to search for one or more combinatorial objects having a desired property out of a set of $M$ objects. Again, $M$ is not necessarily a power of $2$. Typically, one would encode each solution using $n$ bits, for sufficiently large $n$, and perform a Grover search over the space of $N = 2^n$ binary strings (not marking the binary strings which do not correspond to any combinatorial object). Using the indexing unitary however, the search space can be cut down from $N$ to $M$. As per \citet{brassard2000}, the Grover iteration over a set of size $M$ takes the form
	\begin{equation}
	\hat{Q} = -\hat{\mathcal{F}}_M \hat{S}_0 \hat{\mathcal{F}}_M^{-1} \hat{S}_f \, ,
	\end{equation}
	where $\hat{S}_f$ conditionally negates the amplitude of objects meeting the search criteria and $\hat{S}_0$ conditionally negates the amplitude of the all-zero state. The only modification required is $\hat{S}_f \mapsto \hat{S}_{\#} \hat{S}_f \hat{U}_{\#}^\dag$. This un-indexes the integers to their corresponding combinatorial object, so each object can be meaningfully interpreted and marked according to some combinatorial criteria and then be mapped back to an integer index. Thus, the modified Grover iteration is
	\begin{equation}
	Q' = -\hat{\mathcal{F}}_M \hat{S}_0 \hat{\mathcal{F}}_M^{-1} \hat{U}_{\#} \hat{S}_f \hat{U}_{\#}^\dag \, .
	\end{equation}
	Quantum search over a set of $M$ combinatorial objects with an associated indexing function can be performed in $\mathcal{O}(\sqrt{M/k})$ for $k$ marked combinatorial objects. In general, this will lead to a constant-factor speedup.
	
	\section{Conclusion}
	
	In this paper we have established a general framework for combinatorial optimisation via highly efficient continuous-time quantum-walk over finite but exponentially large sets of combinatorial objects. We focus on combinatorial families with an associated `indexing algorithm', which efficiently identifies the position of a given combinatorial objects amongst all objects of the same size. Examples of combinatorial families with associated indexing functions include combinations, permutations, partitions, and lattice walks under a variety of constraints. Using a quantum indexing unitary, the binary representation of the objects can be mapped to a smaller and simpler canonical subspace to allow straightforward implementation of the CTQW. 
	
	This approach is particularly beneficial for use as an quantum walk-based approximate optimisation scheme, to optimise over nontrivial combinatorial domains. The proposed efficient quantum algorithm requires that the choice of graph connecting the combinatorial objects is circulant. The variational parameters in this case are the quantum walk times and the quality phase factors. With this in mind, a specific benefit of our approach is that the size and design of the quantum circuit is completely independent of these parameters, so the circuit does not need to be re-compiled each time these parameters are updated. In addition, by choosing a symmetric graph the optimisation algorithm is not biased towards any solution over another. In this paper we also briefly discuss the relevance to Grover search over combinatorial domains. 
	
	Furthermore, each $\hat{U}_W$ in the QWOA state evolution does not need to remain the same operator. One can consider different connectivities amongst combinatorial objects for each quantum walk. For example, it may be beneficial to start with an initial quantum walk that is highly connected (c.f. complete graph) and decrease the inter-solution connectivity each time, ending with a quantum walk over the objects connected as a cycle graph. This approach may be beneficial to `hone in' on a high-quality solution through a systematically varying $\hat{U}_W$; we leave this to future work.
	
	Finally, it is worth noting that this approach does not work if the combinatorial domain cannot be efficiently indexed, or if its size cannot be efficiently determined. For example, independent sets do not have an efficient indexing function; even the problem of counting the number of independent sets of a graph belongs to the complexity class \#P.


\begin{thebibliography}{46}%
\makeatletter
\providecommand \@ifxundefined [1]{%
 \@ifx{#1\undefined}
}%
\providecommand \@ifnum [1]{%
 \ifnum #1\expandafter \@firstoftwo
 \else \expandafter \@secondoftwo
 \fi
}%
\providecommand \@ifx [1]{%
 \ifx #1\expandafter \@firstoftwo
 \else \expandafter \@secondoftwo
 \fi
}%
\providecommand \natexlab [1]{#1}%
\providecommand \enquote  [1]{``#1''}%
\providecommand \bibnamefont  [1]{#1}%
\providecommand \bibfnamefont [1]{#1}%
\providecommand \citenamefont [1]{#1}%
\providecommand \href@noop [0]{\@secondoftwo}%
\providecommand \href [0]{\begingroup \@sanitize@url \@href}%
\providecommand \@href[1]{\@@startlink{#1}\@@href}%
\providecommand \@@href[1]{\endgroup#1\@@endlink}%
\providecommand \@sanitize@url [0]{\catcode `\\12\catcode `\$12\catcode
  `\&12\catcode `\#12\catcode `\^12\catcode `\_12\catcode `\%12\relax}%
\providecommand \@@startlink[1]{}%
\providecommand \@@endlink[0]{}%
\providecommand \url  [0]{\begingroup\@sanitize@url \@url }%
\providecommand \@url [1]{\endgroup\@href {#1}{\urlprefix }}%
\providecommand \urlprefix  [0]{URL }%
\providecommand \Eprint [0]{\href }%
\providecommand \doibase [0]{http://dx.doi.org/}%
\providecommand \selectlanguage [0]{\@gobble}%
\providecommand \bibinfo  [0]{\@secondoftwo}%
\providecommand \bibfield  [0]{\@secondoftwo}%
\providecommand \translation [1]{[#1]}%
\providecommand \BibitemOpen [0]{}%
\providecommand \bibitemStop [0]{}%
\providecommand \bibitemNoStop [0]{.\EOS\space}%
\providecommand \EOS [0]{\spacefactor3000\relax}%
\providecommand \BibitemShut  [1]{\csname bibitem#1\endcsname}%
\let\auto@bib@innerbib\@empty
\bibitem [{\citenamefont {Korte}\ and\ \citenamefont
  {Vygen}(2007)}]{Korte:2007}%
  \BibitemOpen
  \bibfield  {author} {\bibinfo {author} {\bibfnamefont {B.}~\bibnamefont
  {Korte}}\ and\ \bibinfo {author} {\bibfnamefont {J.}~\bibnamefont {Vygen}},\
  }\href {\doibase 10.1007/3-540-29297-7} {\emph {\bibinfo {title}
  {Combinatorial Optimization: Theory and Algorithms}}},\ \bibinfo {edition}
  {4th}\ ed.\ (\bibinfo  {publisher} {Springer-Verlag Berlin Heidelberg},\
  \bibinfo {year} {2007})\BibitemShut {NoStop}%
\bibitem [{\citenamefont {{Grover}}(1999)}]{Grover1999}%
  \BibitemOpen
  \bibfield  {author} {\bibinfo {author} {\bibfnamefont {L.~K.}\ \bibnamefont
  {{Grover}}},\ }\href {\doibase 10.1016/S0960-0779(98)00217-3} {\bibfield
  {journal} {\bibinfo  {journal} {Chaos Solitons and Fractals}\ }\textbf
  {\bibinfo {volume} {10}},\ \bibinfo {pages} {1695} (\bibinfo {year}
  {1999})}\BibitemShut {NoStop}%
\bibitem [{\citenamefont {Brassard}\ \emph {et~al.}(2002)\citenamefont
  {Brassard}, \citenamefont {H{\o}yer}, \citenamefont {Mosca},\ and\
  \citenamefont {Tapp}}]{brassard2000}%
  \BibitemOpen
  \bibfield  {author} {\bibinfo {author} {\bibfnamefont {G.}~\bibnamefont
  {Brassard}}, \bibinfo {author} {\bibfnamefont {P.}~\bibnamefont {H{\o}yer}},
  \bibinfo {author} {\bibfnamefont {M.}~\bibnamefont {Mosca}}, \ and\ \bibinfo
  {author} {\bibfnamefont {A.}~\bibnamefont {Tapp}},\ }in\ \href {\doibase
  10.1090/conm/305/05215} {\emph {\bibinfo {booktitle} {Quantum computation and
  information ({W}ashington, {DC}, 2000)}}},\ \bibinfo {series} {Contemp.
  Math.}, Vol.\ \bibinfo {volume} {305}\ (\bibinfo  {publisher} {Amer. Math.
  Soc., Providence, RI},\ \bibinfo {year} {2002})\ pp.\ \bibinfo {pages}
  {53--74}\BibitemShut {NoStop}%
\bibitem [{\citenamefont {{Farhi}}\ \emph {et~al.}(2014)\citenamefont
  {{Farhi}}, \citenamefont {{Goldstone}},\ and\ \citenamefont
  {{Gutmann}}}]{Farhi2014}%
  \BibitemOpen
  \bibfield  {author} {\bibinfo {author} {\bibfnamefont {E.}~\bibnamefont
  {{Farhi}}}, \bibinfo {author} {\bibfnamefont {J.}~\bibnamefont
  {{Goldstone}}}, \ and\ \bibinfo {author} {\bibfnamefont {S.}~\bibnamefont
  {{Gutmann}}},\ }\href@noop {} {\bibfield  {journal} {\bibinfo  {journal}
  {arXiv e-prints}\ } (\bibinfo {year} {2014})},\ \Eprint
  {http://arxiv.org/abs/1411.4028} {arXiv:1411.4028 [quant-ph]} \BibitemShut
  {NoStop}%
\bibitem [{\citenamefont {Marsh}\ and\ \citenamefont {Wang}(2019)}]{Marsh2019}%
  \BibitemOpen
  \bibfield  {author} {\bibinfo {author} {\bibfnamefont {S.}~\bibnamefont
  {Marsh}}\ and\ \bibinfo {author} {\bibfnamefont {J.~B.}\ \bibnamefont
  {Wang}},\ }\href {\doibase 10.1007/s11128-019-2171-3} {\bibfield  {journal}
  {\bibinfo  {journal} {Quantum Information Processing}\ }\textbf {\bibinfo
  {volume} {18}},\ \bibinfo {pages} {61} (\bibinfo {year} {2019})}\BibitemShut
  {NoStop}%
\bibitem [{\citenamefont {Farhi}\ and\ \citenamefont
  {Gutmann}(1998)}]{Farhi1998}%
  \BibitemOpen
  \bibfield  {author} {\bibinfo {author} {\bibfnamefont {E.}~\bibnamefont
  {Farhi}}\ and\ \bibinfo {author} {\bibfnamefont {S.}~\bibnamefont
  {Gutmann}},\ }\href {\doibase 10.1103/PhysRevA.58.915} {\bibfield  {journal}
  {\bibinfo  {journal} {Physical Review A (3)}\ }\textbf {\bibinfo {volume}
  {58}},\ \bibinfo {pages} {915} (\bibinfo {year} {1998})}\BibitemShut
  {NoStop}%
\bibitem [{\citenamefont {Kempe}(2003)}]{Kempe2003}%
  \BibitemOpen
  \bibfield  {author} {\bibinfo {author} {\bibfnamefont {J.}~\bibnamefont
  {Kempe}},\ }\href {\doibase 10.1080/00107151031000110776} {\bibfield
  {journal} {\bibinfo  {journal} {Contemporary Physics}\ }\textbf {\bibinfo
  {volume} {44}},\ \bibinfo {pages} {307} (\bibinfo {year} {2003})}\BibitemShut
  {NoStop}%
\bibitem [{\citenamefont {Childs}\ \emph {et~al.}(2013)\citenamefont {Childs},
  \citenamefont {Gosset},\ and\ \citenamefont {Webb}}]{Childs2013}%
  \BibitemOpen
  \bibfield  {author} {\bibinfo {author} {\bibfnamefont {A.~M.}\ \bibnamefont
  {Childs}}, \bibinfo {author} {\bibfnamefont {D.}~\bibnamefont {Gosset}}, \
  and\ \bibinfo {author} {\bibfnamefont {Z.}~\bibnamefont {Webb}},\ }\href
  {\doibase 10.1126/science.1229957} {\bibfield  {journal} {\bibinfo  {journal}
  {Science}\ }\textbf {\bibinfo {volume} {339}},\ \bibinfo {pages} {791}
  (\bibinfo {year} {2013})}\BibitemShut {NoStop}%
\bibitem [{\citenamefont {Manouchehri}\ and\ \citenamefont
  {Wang}(2014)}]{QWBook}%
  \BibitemOpen
  \bibfield  {author} {\bibinfo {author} {\bibfnamefont {K.}~\bibnamefont
  {Manouchehri}}\ and\ \bibinfo {author} {\bibfnamefont {J.}~\bibnamefont
  {Wang}},\ }\href {\doibase 10.1007/978-3-642-36014-5} {\emph {\bibinfo
  {title} {Physical implementation of quantum walks}}},\ Quantum Science and
  Technology\ (\bibinfo  {publisher} {Springer, Heidelberg},\ \bibinfo {year}
  {2014})\BibitemShut {NoStop}%
\bibitem [{\citenamefont {Tang}\ \emph {et~al.}(2018)\citenamefont {Tang},
  \citenamefont {Di~Franco}, \citenamefont {Shi}, \citenamefont {He},
  \citenamefont {Feng}, \citenamefont {Gao}, \citenamefont {Sun}, \citenamefont
  {Li}, \citenamefont {Jiao}, \citenamefont {Wang}, \citenamefont {Kim},\ and\
  \citenamefont {Jin}}]{Tang2018}%
  \BibitemOpen
  \bibfield  {author} {\bibinfo {author} {\bibfnamefont {H.}~\bibnamefont
  {Tang}}, \bibinfo {author} {\bibfnamefont {C.}~\bibnamefont {Di~Franco}},
  \bibinfo {author} {\bibfnamefont {Z.-Y.}\ \bibnamefont {Shi}}, \bibinfo
  {author} {\bibfnamefont {T.-S.}\ \bibnamefont {He}}, \bibinfo {author}
  {\bibfnamefont {Z.}~\bibnamefont {Feng}}, \bibinfo {author} {\bibfnamefont
  {J.}~\bibnamefont {Gao}}, \bibinfo {author} {\bibfnamefont {K.}~\bibnamefont
  {Sun}}, \bibinfo {author} {\bibfnamefont {Z.-M.}\ \bibnamefont {Li}},
  \bibinfo {author} {\bibfnamefont {Z.-Q.}\ \bibnamefont {Jiao}}, \bibinfo
  {author} {\bibfnamefont {T.-Y.}\ \bibnamefont {Wang}}, \bibinfo {author}
  {\bibfnamefont {M.~S.}\ \bibnamefont {Kim}}, \ and\ \bibinfo {author}
  {\bibfnamefont {X.-M.}\ \bibnamefont {Jin}},\ }\href {\doibase
  10.1038/s41566-018-0282-5} {\bibfield  {journal} {\bibinfo  {journal} {Nature
  Photonics}\ }\textbf {\bibinfo {volume} {12}},\ \bibinfo {pages} {754}
  (\bibinfo {year} {2018})}\BibitemShut {NoStop}%
\bibitem [{\citenamefont {Childs}\ and\ \citenamefont
  {Goldstone}(2004)}]{Spatialsearch}%
  \BibitemOpen
  \bibfield  {author} {\bibinfo {author} {\bibfnamefont {A.~M.}\ \bibnamefont
  {Childs}}\ and\ \bibinfo {author} {\bibfnamefont {J.}~\bibnamefont
  {Goldstone}},\ }\href {\doibase 10.1103/PhysRevA.70.022314} {\bibfield
  {journal} {\bibinfo  {journal} {Physical Review A}\ }\textbf {\bibinfo
  {volume} {70}},\ \bibinfo {pages} {022314} (\bibinfo {year}
  {2004})}\BibitemShut {NoStop}%
\bibitem [{\citenamefont {Engel}\ \emph {et~al.}(2007)\citenamefont {Engel},
  \citenamefont {Calhoun}, \citenamefont {Read}, \citenamefont {Ahn},
  \citenamefont {Mancal}, \citenamefont {Cheng}, \citenamefont {Blankenship},\
  and\ \citenamefont {Fleming}}]{Engel2007}%
  \BibitemOpen
  \bibfield  {author} {\bibinfo {author} {\bibfnamefont {G.~S.}\ \bibnamefont
  {Engel}}, \bibinfo {author} {\bibfnamefont {T.~R.}\ \bibnamefont {Calhoun}},
  \bibinfo {author} {\bibfnamefont {E.~L.}\ \bibnamefont {Read}}, \bibinfo
  {author} {\bibfnamefont {T.-K.}\ \bibnamefont {Ahn}}, \bibinfo {author}
  {\bibfnamefont {T.}~\bibnamefont {Mancal}}, \bibinfo {author} {\bibfnamefont
  {Y.-C.}\ \bibnamefont {Cheng}}, \bibinfo {author} {\bibfnamefont {R.~E.}\
  \bibnamefont {Blankenship}}, \ and\ \bibinfo {author} {\bibfnamefont {G.~R.}\
  \bibnamefont {Fleming}},\ }\href {\doibase 10.1038/nature05678} {\bibfield
  {journal} {\bibinfo  {journal} {Nature}\ }\textbf {\bibinfo {volume} {446}},\
  \bibinfo {pages} {782} (\bibinfo {year} {2007})}\BibitemShut {NoStop}%
\bibitem [{\citenamefont {Berry}\ and\ \citenamefont {Wang}(2010)}]{Berry2010}%
  \BibitemOpen
  \bibfield  {author} {\bibinfo {author} {\bibfnamefont {S.~D.}\ \bibnamefont
  {Berry}}\ and\ \bibinfo {author} {\bibfnamefont {J.~B.}\ \bibnamefont
  {Wang}},\ }\href {\doibase 10.1103/PhysRevA.82.042333} {\bibfield  {journal}
  {\bibinfo  {journal} {Physical Review A}\ }\textbf {\bibinfo {volume} {82}},\
  \bibinfo {pages} {042333} (\bibinfo {year} {2010})}\BibitemShut {NoStop}%
\bibitem [{\citenamefont {Lloyd}(1996)}]{Lloyd1996}%
  \BibitemOpen
  \bibfield  {author} {\bibinfo {author} {\bibfnamefont {S.}~\bibnamefont
  {Lloyd}},\ }\href {\doibase 10.1126/science.273.5278.1073} {\bibfield
  {journal} {\bibinfo  {journal} {Science}\ }\textbf {\bibinfo {volume}
  {273}},\ \bibinfo {pages} {1073} (\bibinfo {year} {1996})}\BibitemShut
  {NoStop}%
\bibitem [{\citenamefont {Ambainis}(2007)}]{ambainis2007quantum}%
  \BibitemOpen
  \bibfield  {author} {\bibinfo {author} {\bibfnamefont {A.}~\bibnamefont
  {Ambainis}},\ }\href {\doibase 10.1137/S0097539705447311} {\bibfield
  {journal} {\bibinfo  {journal} {SIAM J. Comput.}\ }\textbf {\bibinfo {volume}
  {37}},\ \bibinfo {pages} {210} (\bibinfo {year} {2007})}\BibitemShut
  {NoStop}%
\bibitem [{\citenamefont {Douglas}\ and\ \citenamefont
  {Wang}(2008)}]{Douglas2008}%
  \BibitemOpen
  \bibfield  {author} {\bibinfo {author} {\bibfnamefont {B.~L.}\ \bibnamefont
  {Douglas}}\ and\ \bibinfo {author} {\bibfnamefont {J.~B.}\ \bibnamefont
  {Wang}},\ }\href {\doibase 10.1088/1751-8113/41/7/075303} {\bibfield
  {journal} {\bibinfo  {journal} {Journal of Physics A: Mathematical and
  Theoretical}\ }\textbf {\bibinfo {volume} {41}},\ \bibinfo {pages} {075303}
  (\bibinfo {year} {2008})}\BibitemShut {NoStop}%
\bibitem [{\citenamefont {Berry}\ and\ \citenamefont {Wang}(2011)}]{Berry2011}%
  \BibitemOpen
  \bibfield  {author} {\bibinfo {author} {\bibfnamefont {S.~D.}\ \bibnamefont
  {Berry}}\ and\ \bibinfo {author} {\bibfnamefont {J.~B.}\ \bibnamefont
  {Wang}},\ }\href {\doibase 10.1103/PhysRevA.83.042317} {\bibfield  {journal}
  {\bibinfo  {journal} {Physical Review A}\ }\textbf {\bibinfo {volume} {83}},\
  \bibinfo {pages} {042317} (\bibinfo {year} {2011})}\BibitemShut {NoStop}%
\bibitem [{\citenamefont {Schreiber}\ \emph {et~al.}(2012)\citenamefont
  {Schreiber}, \citenamefont {G{\'a}bris}, \citenamefont {Rohde}, \citenamefont
  {Laiho}, \citenamefont {{\v S}tefa{\v n}{\'a}k}, \citenamefont {Poto{\v
  c}ek}, \citenamefont {Hamilton}, \citenamefont {Jex},\ and\ \citenamefont
  {Silberhorn}}]{Schreiber2012}%
  \BibitemOpen
  \bibfield  {author} {\bibinfo {author} {\bibfnamefont {A.}~\bibnamefont
  {Schreiber}}, \bibinfo {author} {\bibfnamefont {A.}~\bibnamefont
  {G{\'a}bris}}, \bibinfo {author} {\bibfnamefont {P.~P.}\ \bibnamefont
  {Rohde}}, \bibinfo {author} {\bibfnamefont {K.}~\bibnamefont {Laiho}},
  \bibinfo {author} {\bibfnamefont {M.}~\bibnamefont {{\v S}tefa{\v n}{\'a}k}},
  \bibinfo {author} {\bibfnamefont {V.}~\bibnamefont {Poto{\v c}ek}}, \bibinfo
  {author} {\bibfnamefont {C.}~\bibnamefont {Hamilton}}, \bibinfo {author}
  {\bibfnamefont {I.}~\bibnamefont {Jex}}, \ and\ \bibinfo {author}
  {\bibfnamefont {C.}~\bibnamefont {Silberhorn}},\ }\href {\doibase
  10.1126/science.1218448} {\bibfield  {journal} {\bibinfo  {journal}
  {Science}\ }\textbf {\bibinfo {volume} {336}},\ \bibinfo {pages} {55}
  (\bibinfo {year} {2012})}\BibitemShut {NoStop}%
\bibitem [{\citenamefont {Izaac}\ \emph {et~al.}(2017)\citenamefont {Izaac},
  \citenamefont {Wang}, \citenamefont {Abbott},\ and\ \citenamefont
  {Ma}}]{Izaac2017a}%
  \BibitemOpen
  \bibfield  {author} {\bibinfo {author} {\bibfnamefont {J.~A.}\ \bibnamefont
  {Izaac}}, \bibinfo {author} {\bibfnamefont {J.~B.}\ \bibnamefont {Wang}},
  \bibinfo {author} {\bibfnamefont {P.~C.}\ \bibnamefont {Abbott}}, \ and\
  \bibinfo {author} {\bibfnamefont {X.~S.}\ \bibnamefont {Ma}},\ }\href
  {\doibase 10.1103/physreva.96.032305} {\bibfield  {journal} {\bibinfo
  {journal} {Physical Review A}\ }\textbf {\bibinfo {volume} {96}},\ \bibinfo
  {pages} {032305} (\bibinfo {year} {2017})}\BibitemShut {NoStop}%
\bibitem [{\citenamefont {Levi}(2017)}]{Fleet2017}%
  \BibitemOpen
  \bibfield  {author} {\bibinfo {author} {\bibfnamefont {F.}~\bibnamefont
  {Levi}},\ }\href {https://doi.org/10.1038/nphys4291} {\bibfield  {journal}
  {\bibinfo  {journal} {Nature Physics}\ }\textbf {\bibinfo {volume} {13}},\
  \bibinfo {pages} {926} (\bibinfo {year} {2017})}\BibitemShut {NoStop}%
\bibitem [{\citenamefont {Ming}\ \emph {et~al.}(2019)\citenamefont {Ming},
  \citenamefont {Lin}, \citenamefont {Bartlett},\ and\ \citenamefont
  {Zhang}}]{Ming2019}%
  \BibitemOpen
  \bibfield  {author} {\bibinfo {author} {\bibfnamefont {Y.}~\bibnamefont
  {Ming}}, \bibinfo {author} {\bibfnamefont {C.-T.}\ \bibnamefont {Lin}},
  \bibinfo {author} {\bibfnamefont {S.~D.}\ \bibnamefont {Bartlett}}, \ and\
  \bibinfo {author} {\bibfnamefont {W.-W.}\ \bibnamefont {Zhang}},\ }\href
  {\doibase 10.1038/s41524-019-0224-x} {\bibfield  {journal} {\bibinfo
  {journal} {npj Computational Materials}\ }\textbf {\bibinfo {volume} {5}},\
  \bibinfo {pages} {88} (\bibinfo {year} {2019})}\BibitemShut {NoStop}%
\bibitem [{\citenamefont {Tai}\ \emph {et~al.}(2017)\citenamefont {Tai},
  \citenamefont {Lukin}, \citenamefont {Rispoli}, \citenamefont {Schittko},
  \citenamefont {Menke}, \citenamefont {Borgnia}, \citenamefont {Preiss},
  \citenamefont {Grusdt}, \citenamefont {Kaufman},\ and\ \citenamefont
  {Greiner}}]{Tai2017}%
  \BibitemOpen
  \bibfield  {author} {\bibinfo {author} {\bibfnamefont {M.~E.}\ \bibnamefont
  {Tai}}, \bibinfo {author} {\bibfnamefont {A.}~\bibnamefont {Lukin}}, \bibinfo
  {author} {\bibfnamefont {M.}~\bibnamefont {Rispoli}}, \bibinfo {author}
  {\bibfnamefont {R.}~\bibnamefont {Schittko}}, \bibinfo {author}
  {\bibfnamefont {T.}~\bibnamefont {Menke}}, \bibinfo {author} {\bibfnamefont
  {D.}~\bibnamefont {Borgnia}}, \bibinfo {author} {\bibfnamefont {P.~M.}\
  \bibnamefont {Preiss}}, \bibinfo {author} {\bibfnamefont {F.}~\bibnamefont
  {Grusdt}}, \bibinfo {author} {\bibfnamefont {A.~M.}\ \bibnamefont {Kaufman}},
  \ and\ \bibinfo {author} {\bibfnamefont {M.}~\bibnamefont {Greiner}},\ }\href
  {https://doi.org/10.1038/nature22811} {\bibfield  {journal} {\bibinfo
  {journal} {Nature}\ }\textbf {\bibinfo {volume} {546}},\ \bibinfo {pages}
  {519} (\bibinfo {year} {2017})}\BibitemShut {NoStop}%
\bibitem [{\citenamefont {Harris}\ \emph {et~al.}(2017)\citenamefont {Harris},
  \citenamefont {Steinbrecher}, \citenamefont {Prabhu}, \citenamefont {Lahini},
  \citenamefont {Mower}, \citenamefont {Bunandar}, \citenamefont {Chen},
  \citenamefont {Wong}, \citenamefont {Baehr-Jones}, \citenamefont {Hochberg},
  \citenamefont {Lloyd},\ and\ \citenamefont {Englund}}]{Harris2017}%
  \BibitemOpen
  \bibfield  {author} {\bibinfo {author} {\bibfnamefont {N.~C.}\ \bibnamefont
  {Harris}}, \bibinfo {author} {\bibfnamefont {G.~R.}\ \bibnamefont
  {Steinbrecher}}, \bibinfo {author} {\bibfnamefont {M.}~\bibnamefont
  {Prabhu}}, \bibinfo {author} {\bibfnamefont {Y.}~\bibnamefont {Lahini}},
  \bibinfo {author} {\bibfnamefont {J.}~\bibnamefont {Mower}}, \bibinfo
  {author} {\bibfnamefont {D.}~\bibnamefont {Bunandar}}, \bibinfo {author}
  {\bibfnamefont {C.}~\bibnamefont {Chen}}, \bibinfo {author} {\bibfnamefont
  {F.~N.~C.}\ \bibnamefont {Wong}}, \bibinfo {author} {\bibfnamefont
  {T.}~\bibnamefont {Baehr-Jones}}, \bibinfo {author} {\bibfnamefont
  {M.}~\bibnamefont {Hochberg}}, \bibinfo {author} {\bibfnamefont
  {S.}~\bibnamefont {Lloyd}}, \ and\ \bibinfo {author} {\bibfnamefont
  {D.}~\bibnamefont {Englund}},\ }\href
  {https://doi.org/10.1038/nphoton.2017.95} {\bibfield  {journal} {\bibinfo
  {journal} {Nature Photonics}\ }\textbf {\bibinfo {volume} {11}},\ \bibinfo
  {pages} {447} (\bibinfo {year} {2017})}\BibitemShut {NoStop}%
\bibitem [{\citenamefont {Yan}\ \emph {et~al.}(2019)\citenamefont {Yan},
  \citenamefont {Zhang}, \citenamefont {Gong}, \citenamefont {Wu},
  \citenamefont {Zheng}, \citenamefont {Li}, \citenamefont {Wang},
  \citenamefont {Liang}, \citenamefont {Lin}, \citenamefont {Xu}, \citenamefont
  {Guo}, \citenamefont {Sun}, \citenamefont {Peng}, \citenamefont {Xia},
  \citenamefont {Deng}, \citenamefont {Rong}, \citenamefont {You},
  \citenamefont {Nori}, \citenamefont {Fan}, \citenamefont {Zhu},\ and\
  \citenamefont {Pan}}]{Yan753}%
  \BibitemOpen
  \bibfield  {author} {\bibinfo {author} {\bibfnamefont {Z.}~\bibnamefont
  {Yan}}, \bibinfo {author} {\bibfnamefont {Y.-R.}\ \bibnamefont {Zhang}},
  \bibinfo {author} {\bibfnamefont {M.}~\bibnamefont {Gong}}, \bibinfo {author}
  {\bibfnamefont {Y.}~\bibnamefont {Wu}}, \bibinfo {author} {\bibfnamefont
  {Y.}~\bibnamefont {Zheng}}, \bibinfo {author} {\bibfnamefont
  {S.}~\bibnamefont {Li}}, \bibinfo {author} {\bibfnamefont {C.}~\bibnamefont
  {Wang}}, \bibinfo {author} {\bibfnamefont {F.}~\bibnamefont {Liang}},
  \bibinfo {author} {\bibfnamefont {J.}~\bibnamefont {Lin}}, \bibinfo {author}
  {\bibfnamefont {Y.}~\bibnamefont {Xu}}, \bibinfo {author} {\bibfnamefont
  {C.}~\bibnamefont {Guo}}, \bibinfo {author} {\bibfnamefont {L.}~\bibnamefont
  {Sun}}, \bibinfo {author} {\bibfnamefont {C.-Z.}\ \bibnamefont {Peng}},
  \bibinfo {author} {\bibfnamefont {K.}~\bibnamefont {Xia}}, \bibinfo {author}
  {\bibfnamefont {H.}~\bibnamefont {Deng}}, \bibinfo {author} {\bibfnamefont
  {H.}~\bibnamefont {Rong}}, \bibinfo {author} {\bibfnamefont {J.~Q.}\
  \bibnamefont {You}}, \bibinfo {author} {\bibfnamefont {F.}~\bibnamefont
  {Nori}}, \bibinfo {author} {\bibfnamefont {H.}~\bibnamefont {Fan}}, \bibinfo
  {author} {\bibfnamefont {X.}~\bibnamefont {Zhu}}, \ and\ \bibinfo {author}
  {\bibfnamefont {J.-W.}\ \bibnamefont {Pan}},\ }\href {\doibase
  10.1126/science.aaw1611} {\bibfield  {journal} {\bibinfo  {journal}
  {Science}\ }\textbf {\bibinfo {volume} {364}},\ \bibinfo {pages} {753}
  (\bibinfo {year} {2019})}\BibitemShut {NoStop}%
\bibitem [{\citenamefont {Miri}\ and\ \citenamefont
  {Al{\`u}}(2019)}]{Mirieaar7709}%
  \BibitemOpen
  \bibfield  {author} {\bibinfo {author} {\bibfnamefont {M.-A.}\ \bibnamefont
  {Miri}}\ and\ \bibinfo {author} {\bibfnamefont {A.}~\bibnamefont {Al{\`u}}},\
  }\href {\doibase 10.1126/science.aar7709} {\bibfield  {journal} {\bibinfo
  {journal} {Science}\ }\textbf {\bibinfo {volume} {363}},\ \bibinfo {pages}
  {7709} (\bibinfo {year} {2019})}\BibitemShut {NoStop}%
\bibitem [{\citenamefont {Morley}\ \emph {et~al.}(2019)\citenamefont {Morley},
  \citenamefont {Chancellor}, \citenamefont {Bose},\ and\ \citenamefont
  {Kendon}}]{Morley2019}%
  \BibitemOpen
  \bibfield  {author} {\bibinfo {author} {\bibfnamefont {J.~G.}\ \bibnamefont
  {Morley}}, \bibinfo {author} {\bibfnamefont {N.}~\bibnamefont {Chancellor}},
  \bibinfo {author} {\bibfnamefont {S.}~\bibnamefont {Bose}}, \ and\ \bibinfo
  {author} {\bibfnamefont {V.}~\bibnamefont {Kendon}},\ }\href {\doibase
  10.1103/PhysRevA.99.022339} {\bibfield  {journal} {\bibinfo  {journal} {Phys.
  Rev. A}\ }\textbf {\bibinfo {volume} {99}},\ \bibinfo {pages} {022339}
  (\bibinfo {year} {2019})}\BibitemShut {NoStop}%
\bibitem [{\citenamefont {Callison}\ \emph {et~al.}(2019)\citenamefont
  {Callison}, \citenamefont {Chancellor}, \citenamefont {Mintert},\ and\
  \citenamefont {Kendon}}]{Callison2019}%
  \BibitemOpen
  \bibfield  {author} {\bibinfo {author} {\bibfnamefont {A.}~\bibnamefont
  {Callison}}, \bibinfo {author} {\bibfnamefont {N.}~\bibnamefont
  {Chancellor}}, \bibinfo {author} {\bibfnamefont {F.}~\bibnamefont {Mintert}},
  \ and\ \bibinfo {author} {\bibfnamefont {V.}~\bibnamefont {Kendon}},\ }\href
  {\doibase 10.1088/1367-2630/ab5ca2} {\bibfield  {journal} {\bibinfo
  {journal} {New Journal of Physics}\ }\textbf {\bibinfo {volume} {21}},\
  \bibinfo {pages} {123022} (\bibinfo {year} {2019})}\BibitemShut {NoStop}%
\bibitem [{\citenamefont {Childs}(2004)}]{childs2004quantum}%
  \BibitemOpen
  \bibfield  {author} {\bibinfo {author} {\bibfnamefont {A.~M.}\ \bibnamefont
  {Childs}},\ }\emph {\bibinfo {title} {Quantum information processing in
  continuous time}},\ \href {http://hdl.handle.net/1721.1/16663} {Ph.D.
  thesis},\ \bibinfo  {school} {Massachusetts Institute of Technology}
  (\bibinfo {year} {2004})\BibitemShut {NoStop}%
\bibitem [{\citenamefont {Hadfield}\ \emph {et~al.}(2019)\citenamefont
  {Hadfield}, \citenamefont {Wang}, \citenamefont {O'Gorman}, \citenamefont
  {Rieffel}, \citenamefont {Venturelli},\ and\ \citenamefont
  {Biswas}}]{Hadfield2019}%
  \BibitemOpen
  \bibfield  {author} {\bibinfo {author} {\bibfnamefont {S.}~\bibnamefont
  {Hadfield}}, \bibinfo {author} {\bibfnamefont {Z.}~\bibnamefont {Wang}},
  \bibinfo {author} {\bibfnamefont {B.}~\bibnamefont {O'Gorman}}, \bibinfo
  {author} {\bibfnamefont {E.~G.}\ \bibnamefont {Rieffel}}, \bibinfo {author}
  {\bibfnamefont {D.}~\bibnamefont {Venturelli}}, \ and\ \bibinfo {author}
  {\bibfnamefont {R.}~\bibnamefont {Biswas}},\ }\href {\doibase
  10.3390/a12020034} {\bibfield  {journal} {\bibinfo  {journal} {Algorithms}\
  }\textbf {\bibinfo {volume} {12}},\ \bibinfo {pages} {34} (\bibinfo {year}
  {2019})}\BibitemShut {NoStop}%
\bibitem [{\citenamefont {{Wang}}\ \emph {et~al.}(2019)\citenamefont {{Wang}},
  \citenamefont {{Rubin}}, \citenamefont {{Dominy}},\ and\ \citenamefont
  {{Rieffel}}}]{XYMixers2019}%
  \BibitemOpen
  \bibfield  {author} {\bibinfo {author} {\bibfnamefont {Z.}~\bibnamefont
  {{Wang}}}, \bibinfo {author} {\bibfnamefont {N.~C.}\ \bibnamefont {{Rubin}}},
  \bibinfo {author} {\bibfnamefont {J.~M.}\ \bibnamefont {{Dominy}}}, \ and\
  \bibinfo {author} {\bibfnamefont {E.~G.}\ \bibnamefont {{Rieffel}}},\
  }\href@noop {} {\bibfield  {journal} {\bibinfo  {journal} {arXiv e-prints}\ }
  (\bibinfo {year} {2019})},\ \Eprint {http://arxiv.org/abs/1904.09314}
  {arXiv:1904.09314 [quant-ph]} \BibitemShut {NoStop}%
\bibitem [{\citenamefont {Qiang}\ \emph {et~al.}(2016)\citenamefont {Qiang},
  \citenamefont {Loke}, \citenamefont {Montanaro}, \citenamefont
  {Aungskunsiri}, \citenamefont {Zhou}, \citenamefont {O'Brien}, \citenamefont
  {Wang},\ and\ \citenamefont {Matthews}}]{Qiang2016}%
  \BibitemOpen
  \bibfield  {author} {\bibinfo {author} {\bibfnamefont {X.}~\bibnamefont
  {Qiang}}, \bibinfo {author} {\bibfnamefont {T.}~\bibnamefont {Loke}},
  \bibinfo {author} {\bibfnamefont {A.}~\bibnamefont {Montanaro}}, \bibinfo
  {author} {\bibfnamefont {K.}~\bibnamefont {Aungskunsiri}}, \bibinfo {author}
  {\bibfnamefont {X.}~\bibnamefont {Zhou}}, \bibinfo {author} {\bibfnamefont
  {J.~L.}\ \bibnamefont {O'Brien}}, \bibinfo {author} {\bibfnamefont {J.~B.}\
  \bibnamefont {Wang}}, \ and\ \bibinfo {author} {\bibfnamefont {J.~C.~F.}\
  \bibnamefont {Matthews}},\ }\href {\doibase 10.1038/ncomms11511} {\bibfield
  {journal} {\bibinfo  {journal} {Nature Communications}\ }\textbf {\bibinfo
  {volume} {7}},\ \bibinfo {pages} {11511} (\bibinfo {year}
  {2016})}\BibitemShut {NoStop}%
\bibitem [{\citenamefont {Zhou}\ and\ \citenamefont {Wang}(2017)}]{Zhou2017}%
  \BibitemOpen
  \bibfield  {author} {\bibinfo {author} {\bibfnamefont {S.~S.}\ \bibnamefont
  {Zhou}}\ and\ \bibinfo {author} {\bibfnamefont {J.~B.}\ \bibnamefont
  {Wang}},\ }\href {\doibase 10.1098/rsos.160906} {\bibfield  {journal}
  {\bibinfo  {journal} {R. Soc. Open Sci.}\ }\textbf {\bibinfo {volume} {4}},\
  \bibinfo {pages} {160906, 12} (\bibinfo {year} {2017})}\BibitemShut {NoStop}%
\bibitem [{\citenamefont {Zhou}\ \emph {et~al.}(2017)\citenamefont {Zhou},
  \citenamefont {Loke}, \citenamefont {Izaac},\ and\ \citenamefont
  {Wang}}]{Zhou2017qft}%
  \BibitemOpen
  \bibfield  {author} {\bibinfo {author} {\bibfnamefont {S.~S.}\ \bibnamefont
  {Zhou}}, \bibinfo {author} {\bibfnamefont {T.}~\bibnamefont {Loke}}, \bibinfo
  {author} {\bibfnamefont {J.~A.}\ \bibnamefont {Izaac}}, \ and\ \bibinfo
  {author} {\bibfnamefont {J.~B.}\ \bibnamefont {Wang}},\ }\href {\doibase
  10.1007/s11128-017-1515-0} {\bibfield  {journal} {\bibinfo  {journal}
  {Quantum Information Processing}\ }\textbf {\bibinfo {volume} {16}},\
  \bibinfo {pages} {82} (\bibinfo {year} {2017})}\BibitemShut {NoStop}%
\bibitem [{\citenamefont {{Qiang}}\ \emph {et~al.}(2018)\citenamefont
  {{Qiang}}, \citenamefont {{Zhou}}, \citenamefont {{Wang}}, \citenamefont
  {{Wilkes}}, \citenamefont {{Loke}}, \citenamefont {{O’Gara}}, \citenamefont
  {{Kling}}, \citenamefont {{Marshall}}, \citenamefont {{Santagati}},
  \citenamefont {{Ralph}}, \citenamefont {{Wang}}, \citenamefont {{O’Brien}},
  \citenamefont {{Thompson}},\ and\ \citenamefont {{Matthews}}}]{Qiang2018}%
  \BibitemOpen
  \bibfield  {author} {\bibinfo {author} {\bibfnamefont {X.}~\bibnamefont
  {{Qiang}}}, \bibinfo {author} {\bibfnamefont {X.~Q.}\ \bibnamefont {{Zhou}}},
  \bibinfo {author} {\bibfnamefont {J.~W.}\ \bibnamefont {{Wang}}}, \bibinfo
  {author} {\bibfnamefont {C.~M.}\ \bibnamefont {{Wilkes}}}, \bibinfo {author}
  {\bibfnamefont {T.}~\bibnamefont {{Loke}}}, \bibinfo {author} {\bibfnamefont
  {S.}~\bibnamefont {{O’Gara}}}, \bibinfo {author} {\bibfnamefont
  {L.}~\bibnamefont {{Kling}}}, \bibinfo {author} {\bibfnamefont {G.~D.}\
  \bibnamefont {{Marshall}}}, \bibinfo {author} {\bibfnamefont
  {R.}~\bibnamefont {{Santagati}}}, \bibinfo {author} {\bibfnamefont {T.~C.}\
  \bibnamefont {{Ralph}}}, \bibinfo {author} {\bibfnamefont {J.~B.}\
  \bibnamefont {{Wang}}}, \bibinfo {author} {\bibfnamefont {J.~L.}\
  \bibnamefont {{O’Brien}}}, \bibinfo {author} {\bibfnamefont {M.~G.}\
  \bibnamefont {{Thompson}}}, \ and\ \bibinfo {author} {\bibfnamefont
  {J.~C.~F.}\ \bibnamefont {{Matthews}}},\ }\href@noop {} {\bibfield  {journal}
  {\bibinfo  {journal} {Nature Photonics}\ }\textbf {\bibinfo {volume} {12}},\
  \bibinfo {pages} {534} (\bibinfo {year} {2018})}\BibitemShut {NoStop}%
\bibitem [{\citenamefont {{Kitaev}}(1995)}]{Kitaev1995}%
  \BibitemOpen
  \bibfield  {author} {\bibinfo {author} {\bibfnamefont {A.~Y.}\ \bibnamefont
  {{Kitaev}}},\ }\href@noop {} {\bibfield  {journal} {\bibinfo  {journal}
  {arXiv e-prints}\ } (\bibinfo {year} {1995})},\ \Eprint
  {http://arxiv.org/abs/quant-ph/9511026} {arXiv:quant-ph/9511026 [quant-ph]}
  \BibitemShut {NoStop}%
\bibitem [{\citenamefont {Wilf}(1977)}]{WILF1977281}%
  \BibitemOpen
  \bibfield  {author} {\bibinfo {author} {\bibfnamefont {H.~S.}\ \bibnamefont
  {Wilf}},\ }\href {\doibase https://doi.org/10.1016/S0001-8708(77)80046-7}
  {\bibfield  {journal} {\bibinfo  {journal} {Advances in Mathematics}\
  }\textbf {\bibinfo {volume} {24}},\ \bibinfo {pages} {281 } (\bibinfo {year}
  {1977})}\BibitemShut {NoStop}%
\bibitem [{\citenamefont {{Cleve}}\ and\ \citenamefont
  {{Watrous}}(2000)}]{Cleve2000}%
  \BibitemOpen
  \bibfield  {author} {\bibinfo {author} {\bibfnamefont {R.}~\bibnamefont
  {{Cleve}}}\ and\ \bibinfo {author} {\bibfnamefont {J.}~\bibnamefont
  {{Watrous}}},\ }in\ \href {\doibase 10.1109/SFCS.2000.892140} {\emph
  {\bibinfo {booktitle} {Proceedings 41st Annual Symposium on Foundations of
  Computer Science}}}\ (\bibinfo {year} {2000})\ pp.\ \bibinfo {pages}
  {526--536}\BibitemShut {NoStop}%
\bibitem [{\citenamefont {Bravyi}\ \emph {et~al.}(2018)\citenamefont {Bravyi},
  \citenamefont {Gosset},\ and\ \citenamefont {K{\"o}nig}}]{Bravyi308}%
  \BibitemOpen
  \bibfield  {author} {\bibinfo {author} {\bibfnamefont {S.}~\bibnamefont
  {Bravyi}}, \bibinfo {author} {\bibfnamefont {D.}~\bibnamefont {Gosset}}, \
  and\ \bibinfo {author} {\bibfnamefont {R.}~\bibnamefont {K{\"o}nig}},\ }\href
  {\doibase 10.1126/science.aar3106} {\bibfield  {journal} {\bibinfo  {journal}
  {Science}\ }\textbf {\bibinfo {volume} {362}},\ \bibinfo {pages} {308}
  (\bibinfo {year} {2018})}\BibitemShut {NoStop}%
\bibitem [{\citenamefont {{OEIS Foundation Inc.}}(2019)}]{oeis}%
  \BibitemOpen
  \bibfield  {author} {\bibinfo {author} {\bibnamefont {{OEIS Foundation
  Inc.}}},\ }\href@noop {} {\enquote {\bibinfo {title} {{The On-Line
  Encyclopedia of Integer Sequences}},}\ }\bibinfo {howpublished}
  {\url{http://oeis.org}} (\bibinfo {year} {2019})\BibitemShut {NoStop}%
\bibitem [{\citenamefont {Lalou}\ \emph {et~al.}(2018)\citenamefont {Lalou},
  \citenamefont {Tahraoui},\ and\ \citenamefont {Kheddouci}}]{LALOU201892}%
  \BibitemOpen
  \bibfield  {author} {\bibinfo {author} {\bibfnamefont {M.}~\bibnamefont
  {Lalou}}, \bibinfo {author} {\bibfnamefont {M.~A.}\ \bibnamefont {Tahraoui}},
  \ and\ \bibinfo {author} {\bibfnamefont {H.}~\bibnamefont {Kheddouci}},\
  }\href {\doibase https://doi.org/10.1016/j.cosrev.2018.02.002} {\bibfield
  {journal} {\bibinfo  {journal} {Computer Science Review}\ }\textbf {\bibinfo
  {volume} {28}},\ \bibinfo {pages} {92 } (\bibinfo {year} {2018})}\BibitemShut
  {NoStop}%
\bibitem [{\citenamefont {Myrvold}\ and\ \citenamefont
  {Ruskey}(2001)}]{MYRVOLD2001281}%
  \BibitemOpen
  \bibfield  {author} {\bibinfo {author} {\bibfnamefont {W.}~\bibnamefont
  {Myrvold}}\ and\ \bibinfo {author} {\bibfnamefont {F.}~\bibnamefont
  {Ruskey}},\ }\href {\doibase https://doi.org/10.1016/S0020-0190(01)00141-7}
  {\bibfield  {journal} {\bibinfo  {journal} {Information Processing Letters}\
  }\textbf {\bibinfo {volume} {79}},\ \bibinfo {pages} {281 } (\bibinfo {year}
  {2001})}\BibitemShut {NoStop}%
\bibitem [{\citenamefont {{K\'asa}}(2009)}]{Kasa2010}%
  \BibitemOpen
  \bibfield  {author} {\bibinfo {author} {\bibfnamefont {Z.}~\bibnamefont
  {{K\'asa}}},\ }\href {https://zbmath.org/?q=an\%3A1178.68371} {\bibfield
  {journal} {\bibinfo  {journal} {{Acta Univ. Sapientiae, Inform.}}\ }\textbf
  {\bibinfo {volume} {1}},\ \bibinfo {pages} {109} (\bibinfo {year}
  {2009})}\BibitemShut {NoStop}%
\bibitem [{\citenamefont {Loehr}(2011)}]{Loehr2011}%
  \BibitemOpen
  \bibfield  {author} {\bibinfo {author} {\bibfnamefont {N.~A.}\ \bibnamefont
  {Loehr}},\ }\href {\doibase 10.1201/b17376} {\emph {\bibinfo {title}
  {Bijective combinatorics}}},\ Discrete Mathematics and its Applications (Boca
  Raton)\ (\bibinfo  {publisher} {CRC Press, Boca Raton, FL},\ \bibinfo {year}
  {2011})\ pp.\ \bibinfo {pages} {187--188}\BibitemShut {NoStop}%
\bibitem [{\citenamefont {B{\"a}rtschi}\ and\ \citenamefont
  {Eidenbenz}(2019)}]{Bartschi2019}%
  \BibitemOpen
  \bibfield  {author} {\bibinfo {author} {\bibfnamefont {A.}~\bibnamefont
  {B{\"a}rtschi}}\ and\ \bibinfo {author} {\bibfnamefont {S.}~\bibnamefont
  {Eidenbenz}},\ }in\ \href@noop {} {\emph {\bibinfo {booktitle} {Fundamentals
  of Computation Theory}}}\ (\bibinfo  {publisher} {Springer International
  Publishing},\ \bibinfo {address} {Cham},\ \bibinfo {year} {2019})\ pp.\
  \bibinfo {pages} {126--139}\BibitemShut {NoStop}%
\bibitem [{\citenamefont {Chiew}\ \emph {et~al.}(2019)\citenamefont {Chiew},
  \citenamefont {de~Lacy}, \citenamefont {Yu}, \citenamefont {Marsh},\ and\
  \citenamefont {Wang}}]{Chiew2019}%
  \BibitemOpen
  \bibfield  {author} {\bibinfo {author} {\bibfnamefont {M.}~\bibnamefont
  {Chiew}}, \bibinfo {author} {\bibfnamefont {K.}~\bibnamefont {de~Lacy}},
  \bibinfo {author} {\bibfnamefont {C.~H.}\ \bibnamefont {Yu}}, \bibinfo
  {author} {\bibfnamefont {S.}~\bibnamefont {Marsh}}, \ and\ \bibinfo {author}
  {\bibfnamefont {J.~B.}\ \bibnamefont {Wang}},\ }\href {\doibase
  10.1007/s11128-019-2407-2} {\bibfield  {journal} {\bibinfo  {journal}
  {Quantum Information Processing}\ }\textbf {\bibinfo {volume} {18}},\
  \bibinfo {pages} {302} (\bibinfo {year} {2019})}\BibitemShut {NoStop}%
\bibitem [{\citenamefont {Berry}\ \emph {et~al.}(2018)\citenamefont {Berry},
  \citenamefont {Kieferov{\'a}}, \citenamefont {Scherer}, \citenamefont
  {Sanders}, \citenamefont {Low}, \citenamefont {Wiebe}, \citenamefont
  {Gidney},\ and\ \citenamefont {Babbush}}]{Berry2018}%
  \BibitemOpen
  \bibfield  {author} {\bibinfo {author} {\bibfnamefont {D.~W.}\ \bibnamefont
  {Berry}}, \bibinfo {author} {\bibfnamefont {M.}~\bibnamefont
  {Kieferov{\'a}}}, \bibinfo {author} {\bibfnamefont {A.}~\bibnamefont
  {Scherer}}, \bibinfo {author} {\bibfnamefont {Y.~R.}\ \bibnamefont
  {Sanders}}, \bibinfo {author} {\bibfnamefont {G.~H.}\ \bibnamefont {Low}},
  \bibinfo {author} {\bibfnamefont {N.}~\bibnamefont {Wiebe}}, \bibinfo
  {author} {\bibfnamefont {C.}~\bibnamefont {Gidney}}, \ and\ \bibinfo {author}
  {\bibfnamefont {R.}~\bibnamefont {Babbush}},\ }\href {\doibase
  10.1038/s41534-018-0071-5} {\bibfield  {journal} {\bibinfo  {journal} {npj
  Quantum Information}\ }\textbf {\bibinfo {volume} {4}},\ \bibinfo {pages}
  {22} (\bibinfo {year} {2018})}\BibitemShut {NoStop}%
\end{thebibliography}
\end{document}